\documentclass{aa}
\usepackage{txfonts}
\usepackage{graphicx}
\usepackage{caption}
\usepackage{subcaption}
\usepackage{color}
\usepackage{tablefootnote}

\usepackage{tikz}
\usetikzlibrary{arrows,shapes,backgrounds}
\usepackage{hyperref}
\hypersetup{colorlinks=true, citecolor=blue, urlcolor=blue}

\begin{document} 
\titlerunning{}
\authorrunning{Mountrichas et al.}
\titlerunning{Comparison of the star formation of X-ray AGN in eFEDS with star-forming galaxies}

\title{Comparison of the star formation of X-ray selected active galactic nuclei in eFEDS with star-forming galaxies}

\author{G. Mountrichas\inst{1}, V. Buat\inst{2,3}, G. Yang\inst{4,5}, M. Boquien\inst{6}, D. Burgarella\inst{2}, L. Ciesla\inst{2}, K. Malek\inst{7,2}, R. Shirley\inst{8,9}}
          
     \institute {Instituto de Fisica de Cantabria (CSIC-Universidad de Cantabria), Avenida de los Castros, 39005 Santander, Spain
              \email{gmountrichas@gmail.com}
           \and
             Aix Marseille Univ, CNRS, CNES, LAM Marseille, France. 
                \email{ veronique.buat@lam.fr}  
              \and
                 Institut Universitaire de France (IUF)
                \and
                Department of Physics and Astronomy, Texas A\&M University, College Station, TX 77843-4242, USA 
               \and
                George P. and Cynthia Woods Mitchell Institute for Fundamental Physics and Astronomy, Texas A\&M University, College Station, TX 77843-4242, USA 
                \and
                Centro de Astronom\'ia (CITEVA), Universidad de Antofagasta, Avenida Angamos 601, Antofagasta, Chile
          \and
		National Centre for Nuclear Research, ul. Pasteura 7, 02-093, Warszawa, Poland
     \and
	Astronomy Centre, Department of Physics \& Astronomy, University of Southampton, Southampton, SO17 1BJ, UK 
	 \and
    Institute of Astronomy, University of Cambridge, Madingley Road,
  Cambridge CB3 0HA, UK}

\abstract {We use $\sim 1800$ X-ray Active Galactic Nuclei (AGN) in the eROSITA Final Equatorial-Depth Survey (eFEDS), that span over two orders of magnitude in X-ray luminosity, $\rm L_{X,2-10keV}  \approx 10^{43-45}\,ergs^{-1}$, and compare their star-formation rate (SFR) relative to that of non-AGN star-forming systems, at $\rm 0.5<z<1.5$. For that purpose, we compile a reference galaxy catalogue with $\sim 17000$ sources. Both samples have the same photometric coverage, from optical to far-infrared. We construct the spectral energy distributions (SEDs) of all sources and fit them using the CIGALE code, utilizing the same templates and parametric grid for both samples. We account for the mass incompleteness and exclude quiescent systems from both datasets. These allow us to compare the SFR of the two populations in a uniform manner, minimising systematic effects. Based on our analysis, AGN at low and moderate L$_X$ ($\rm L_{X,2-10keV} < 10^{44}\,ergs^{-1}$), have SFR that is lower, or at most, equal to that of star-forming galaxies, in agreement with previous studies. The large number of luminous  X-ray AGN available in the eFEDS dataset, enable us to expand our investigations at higher L$_X$ to test previous, tentative results. At $\rm L_{X,2-10keV} > 10^{44.2}\,ergs^{-1}$, the SFR of AGN appears enhanced, by $\sim 30\%$, compared to that of star-forming sources, for systems with stellar mass, $\rm 10.5 < log\,[M_*(M_\odot)] < 11.5$, confirming indications found in previous studies. The most massive sources $\rm log\,[M_*(M_\odot)] > 11.5$, present a flat SFR$_{norm}$-L$_X$ relation up to $\rm L_{X,2-10keV} \sim 10^{44.5}\,ergs^{-1}$, with SFR similar to that of star-forming galaxies. However, at higher L$_X$ ($\rm L_{X,2-10keV} \sim 10^{45}\,ergs^{-1}$), we find indications that the SFR of these massive AGN hosts may be enhanced compared to that of non-AGN systems.} 

   
\maketitle

\section{Introduction}

It has been more than two decades since a relation was found between the large scale properties of galaxies and the mass of the supermassive black holes (SMBH) that live in their centres \citep[e.g.][]{Magorrian1998, Ferrarese2000, Gebhardt2000, Kormendy2013}. It still is, though, a matter of debate how SMBHs and galaxies interact with each other, despite their difference in physical scale  \citep[e.g.][]{Alexander2012}. 

To shed light on this question it is important  to uncover the mechanisms that drive cold gas onto the SMBH, causing its growth and giving birth to an Active Galactic Nuclei (AGN). Various fuelling processes have been proposed in the literature \citep[for a review see ][]{Alexander2012}, depending on the redshift, the AGN power (X-ray luminosity, L$_X$) and the stellar mass, M$_*$, of the galaxy. Major mergers provide a possible triggering mechanism in the case of luminous AGN \citep[e.g.,][]{Bower2006, Hopkins2008a}, while for lower luminosity AGN alternative processes have been suggested, such as minor mergers and disk instabilities \citep[e.g.,][]{Genzel2008, Ciotti2010}. In these cases, large amounts of gas are driven to the centre of galaxies, fuelling the SMBH and setting off the star-formation (SF) of the host galaxy. Therefore, the same mechanism is responsible for triggering both the growth of SMBH and the growth of the galaxy itself. However, alternative processes have been proposed that activate the SMBH, but are decoupled from the SF of the (host) galaxy. For example, in massive systems, diffuse hot gas can be accreted onto the SMBH without first being cooled onto the galactic plane \citep[e.g.,][]{Fanidakis2013}.  

AGN feedback can also regulate the SF activity either by heating the gas reservoir of the host galaxy \citep[negative feedback, e.g.,][]{DiMatteo2005, Croton2006} or by triggering the SF via e.g., AGN outflows during the gas-rich phase of galaxies \citep[positive feedback, e.g.,][]{Zubovas2013}. Understanding the physical mechanisms that trigger the AGN activity and the interplay between AGN and SF, over a wide range of L$_X$, redshift and M$_*$ is fundamental to understand the galaxy formation and evolution. 

Towards this end, a number of works have studied the SF of galaxies that host AGN as a function of AGN power \citep[e.g.,][]{Lutz2010, Rosario2012, Rovilos2012}. More information can be gained, though, by comparing the SFR of AGN host galaxies with the SFR of non-AGN systems. At low redshifts ($\rm z<0.5$), studies found that  the SFR of AGN is consistent with that of main sequence (MS) star-forming galaxies \citep[e.g.,][]{Santini2012, Shimizu2015, Leslie2016, Shimizu2017}. Caution should be taken, though, when comparing results from different studies, since different works use galaxy control samples that include sources with different properties. For instance, as mentioned above, the SFR of X-ray AGN is consistent with that of star-forming galaxies, but appears higher, compared with a galaxy sample that includes both star-forming and quiescent systems \citep{Santini2012} or with a simple mass-matching control sample \citep[see Sect. 7.4 in ][]{Shimizu2017}. 

Luminous AGN are more rare compared to their lower L$_X$ counterparts and therefore larger cosmic volumes need to be probed to sample them. At higher redshifts, the scarcity of galaxies compelled most X-ray studies to use analytical expressions from the literature to measure the SFR of MS star-forming galaxies \citep[e.g.,][]{Schreiber2015} and compare it with the SFR of AGN \citep[e.g.,][]{Rosario2013, Mullaney2015, Masoura2018, Bernhard2019, Masoura2021}. The majority of these works utilized the SFR$_{norn}$ parameter to quantify this comparison. SFR$_{norm}$ is defined as the ratio of the SFR of AGN dominated systems to the SFR of star-forming galaxies with the same M$_*$ and redshift. Based on their findings, SFR$_{norm}$ is independent of redshift \citep{Mullaney2015}. Moreover, there is a strong dependence of SFR$_{norm}$ with the X-ray luminosity \citep{Masoura2021}, with more luminous AGN to have a narrower SFR$_{norm}$ distribution, shifted to higher values and close to those of MS galaxies compared to their lower L$_X$ counterparts \citep{Bernhard2019}. 

Recently, a number of studies compiled large non-AGN galaxy samples and compared their SFR with that of X-ray AGN.  \cite{Florez2020} used AGN with $\rm L_{X,2-10keV}  > 10^{44}\,ergs^{-1}$ in the Bo$\rm \ddot{o}$tes field and compared their SFR with a large, comparison sample of $\sim 320000$ sources without X-ray AGN. Based on their analysis, the average SFR of galaxies that host luminous AGN is higher by a factor of $\sim 3-10$ compared to sources without AGN, at fixed M$_*$ and redshift. 

\cite{Mountrichas2021c} using luminous X-ray AGN in the Bo$\rm \ddot{o}$tes field ($\rm L_{X,2-10keV}  \sim 10^{43.5-45}\,ergs^{-1}$), demonstrated the importance of using a galaxy control sample to compare the SFR of X-ray with non-AGN systems. Their analysis showed that, utilizing an analytical expression from the literature to calculate the SFR of star-forming MS systems, may introduce systematics that affect the comparison of the SFR of the two populations. Their results indicated that at high L$_X$ ($\rm L_{X,2-10keV} > 2-3\times 10^{44}\,ergs^{-1}$), AGN hosted by galaxies with stellar mass $\rm 10.5 < log\,[M_*(M_\odot)] < 11.5$, have enhanced SFR by $\sim 50\%$ compared to their galaxy reference sample. \cite{Mountrichas2022} used X-ray sources from the {\it{COSMOS-Legacy}} survey and applied the same methodology with \cite{Mountrichas2021c}. Their AGN spanned lower luminosities ($\rm L_{X,2-10keV}  \sim 10^{42.5-44}\,ergs^{-1}$) compared to their X-ray counterparts in Bo$\rm \ddot{o}$tes. Based on their results, low to moderate luminosity AGN have SFR that is lower, or at most equal, to that of MS galaxies. 
 
In this work, we use X-ray AGN from the eROSITA Final Equatorial Depth Survey (eFEDS) field and compare their SFR with a galaxy reference catalogue, within the same spatial volume. The datasets are described in detail in Sect. \ref{sec_data}. We follow the same methodology applied in the previous works of \cite{Mountrichas2021c, Mountrichas2022} which allows us to compare and complement the results, covering in total a luminosity baseline of more than 2.5 orders of magnitude ($\rm L_{X,2-10keV} \sim 10^{42.5-45}\,ergs^{-1}$). Specifically, we construct the spectral energy distributions (SEDs) of all galaxies and fit them, using the CIGALE code \citep{Boquien2019, Yang2020, Yang2022}. The models, parametric grid and quality criteria are described in Sect. \ref{sec_analysis} and are identical to those used in \cite{Mountrichas2021c, Mountrichas2022}, to avoid systematic effects introduced by different templates and parameter space. Our main goal is to use the large number of luminous AGN available in the eFEDS field and examine whether the tentative results presented in \cite{Mountrichas2021c} are confirmed. Our measurements are presented in Sect. \ref{sec_lx_sfr}. 
 
Throughout this work, we assume a flat $\Lambda$CDM cosmology with $H_ 0=70.4$\,km\,s$^{-1}$\,Mpc$^{-1}$ and $\Omega _ M=0.272$ \citep{WMAP7}.

\section{Data}
\label{sec_data}

\subsection{X-ray sample}

In our analysis, we use the X-ray sources observed in the eFEDS field. The catalogue is presented in \cite{Brunner2021}. eROSITA \citep[extended ROentgen Survey with an Imaging Telescope Array;][]{Predehl2021} is the primary instrument on the Spektrum-Roentgen-Gamma (SRG) orbital observatory \citep{Sunyaev2021}. SRG was built to provide a sensitive, wide field-of-view X-ray telescope with improved capabilities compared to those of XMM-Newton and Chandra, the two most sensitive targeting X-ray telescopes in operation. 

The catalogue includes 27910 X-ray sources\footnote{An updated catalogue was released on December 3rd, 2021. This updated version is used in our analysis.}, detected in the $0.2-2.3$\,keV energy band with detection likelihoods $\geq 6$, that corresponds to a flux limit of $\approx 7 \times 10^{-15}$\,erg\,cm$^{-2}\,\rm s{^{-1}}$ in the $0.5-2.0$\,keV energy range \citep{Brunner2021}. About $\sim 3\%$ of the sources are located at the borders of the field, which implies shorter exposure times, stronger vignetting and higher background. These sources are excluded from our analysis ("{\textsc{inArea90}}" flag). \cite{Salvato2021} presented the multiwavelength counterparts and redshifts of the X-ray sources, by identifying their optical counterparts. The DESI Legacy Imaging Survey DR8 \citep[LS8;][]{Dey2019} was used for the counterparts identification, due to its homogeneous coverage of the field and its depth. The catalogue also includes Gaia \citep{gaia2020} and WISE \citep{Lang2014} photometry. Two independent methods were utilized to find the counterparts of the X-ray sources, {\tiny {NWAY}} \citep{Salvato2018b} and {\tiny {ASTROMATCH}} \citep{Ruiz2018}. {\tiny {NWAY}} is based on Bayesian statistics and {\tiny {ASTROMATCH}} on the Maximum Likelihood Ratio \citep{Sutherland_and_Saunders1992}. For $88.4\%$ of the eFEDS point like sources, the two methods point at the same counterpart. Each counterpart is assigned a quality flag, {\textsc{CTP\_quality}}. Counterparts with $\rm {\textsc{CTP\_quality}} \geq 2$ are considered reliable, in the sense that either both methods agree on the counterpart and have assigned a counterpart probability above threshold ($\rm {\textsc{CTP\_quality}} = 4$ for 20873 sources), or both methods agree on the counterpart but one method has assigned a probability above threshold ($\rm {\textsc{{\textsc{CTP\_quality}}}} = 3$, 1379 sources), or there is more than one possible counterparts ($\rm {\textsc{{\textsc{CTP\_quality}}}} = 2$, 2522 sources). Only sources with $\rm {\textsc{{\textsc{CTP\_quality}}}} \geq 2$ are included in our analysis ($24,774/27,910$). We note, however, that sources with  $\rm {\textsc{{\textsc{CTP\_quality}}}} = 2$ represent only $4\%$ of our final X-ray sample and their exclusion from our analysis would not affect our results and conclusions (for the final selection of X-ray sources, see Sect. \ref{sec_analysis}). Sources were, then, classified into Galactic and extragalactic, using a combination of methods and various information \citep[for more details see Sect. 5 in][]{Salvato2021}. 21952 out of the 24,774 X-ray sources are characterised as extragalactic. Galactic sources are rejected from our analysis. 

eFEDS has been observed by a number of spectroscopic surveys, such as GAMA \citep{Baldry2018}, SDSS \citep{Blanton2017} and WiggleZ \citep{Drinkwater2018}. Only sources with secure spectroscopic redshift, {\textit{specz}}, from the parent catalogues were considered in the eFEDS catalogue \citep{Salvato2021}. 6640 sources have reliable {\textit{specz}}. Photometric redshifts, {\textit{photoz}}, were computed for the remaining sources using the LePHARE code \citep{Arnouts1999, Ilbert2006} and following the procedure outlined in e.g., \cite{Salvato2009, Salvato2011, Fotopoulou2012}. These estimates were compared with those using DNNz, a machine learning algorithm that uses exclusively HSC photometry (Nishizawa et al., in prep.). A redshift flag is assigned to each source, CTP\_REDSHIFT\_GRADE. Only sources with $\rm CTP\_REDSHIFT\_GRADE \geq 3$ (26047/27910) are considered in this work. This criterion includes sources with either spectroscopic redshift ($\rm CTP\_REDSHIFT\_GRADE = 5$) or the {\textit{photoz}} estimates of the two methods agree ($\rm CTP\_REDSHIFT\_GRADE = 4 $) or agree within a tolerance level \citep[$\rm CTP\_REDSHIFT\_GRADE = 3$, for more details see Sect. 6.3 of][]{Salvato2021}. We note, that $80\%$ of the X-ray sources in our final sample (see Sect. \ref{sec_analysis}) have $\rm CTP\_REDSHIFT\_GRADE \geq 4$. Furthermore, we restrict our sources to those within the KiDS+VIKING area \citep{Kuijken2019, Hildebrandt2020}. Near-infrared (NIR) photometry outside of this region is shallow which significantly affects the accuracy and reliability of the {\textit{photoz}} calculations \citep[Sect. 6.1 in][]{Salvato2021}. Based on the numbers quoted in Table 7 of \cite{Salvato2021}, the accuracy of {\it{photoz}} within the KiDs area, is $\sigma_{\Delta z/(1+z_{spec})}=0.049$ and the fraction of outliers ($\Delta z/(1+z_{{\textit{specz}}})>0.15$) is $13.9\%$. 10,294/21,952 extragalactic X-ray sources are included in this area. Applying the $\rm {\textsc{{\textsc{CTP\_quality}}}} \geq 2$ and $\rm CTP\_REDSHIFT\_GRADE \geq 3$ criteria, reduces the available number of X-ray sources to 10,098 AGN for our analysis.

\cite{Liu2021} performed a systematic X-ray spectral fitting analysis on all the X-ray systems, providing fluxes and luminosities, among other X-ray properties, for the eFEDS sources. Based on their results only $10\%$ of the sources are X-ray obscured. The power-law slope calculations are described by a Gaussian distribution with mean value and dispersion of $1.94 \pm0.22$. In this work, we use their posterior median, intrinsic (absorption corrected) X-ray fluxes in the $2-10$\,keV energy band.

In our analysis, we measure (host) galaxy properties via SED fitting. In order to get reliable results, it is essential to measure these galaxy properties with the highest possible accuracy. Therefore, we require all X-ray AGN to have available the following photometric bands $u, g, r, i, z, J, H, K, W1, W2, W4$, where W1, W2, W4 are the photometric bands of WISE \citep{Wright2010}, at 3.4\,$\mu m$, 4.6\,$\mu m$ and 22]\,$\mu m$, respectively, and the others are the optical and NIR photometric bands of KiDS/VIKING. These criteria, reduce the X-ray sources to 5921, i.e. $\sim 40\%$ of the 10,098 are rejected. This is due to our requirement for W4 ($35\%$ of the 10098 do not have a W4 measurement). However, W4 is important to fit the mid-infrared (mid-IR) continuum, in particular at $\rm z>1$. We note that, although, we do not apply a requirement for availability of the W3 band (WISE band at 12\,$\mu m$), $83\%$ of the X-ray sources in the final sample (see Sect. \ref{sec_analysis}) have W3 measurement. 

%(KEEP IN MIND that this number will reduce to 4354 after x2 and bayes/best criteria will be applied). 

The X-ray catalogue is also cross-matched with the GAMA-09 photometric catalogue produced by the HELP collaboration \citep{Shirley2019, Shirley2021}, that covers $\sim 35\%$ of the eFEDS area. HELP provides data from 23 extragalactic survey fields, imaged by the {\it{Herschel}} Space Observatory which form the {\it{Herschel}} Extragalactic Legacy Project (HELP). The position of NIR/IRAC sources are then used as prior information to extract sources in the {\it{Herschel}} maps. The XID+ tool \citep{Hurley2017}, developed for this purpose, uses a Bayesian probabilistic framework and works with prior positions. The cross-match between the two catalogues was performed using 1\arcsec radius and the optical coordinates of the counterpart of each X-ray source. About $\sim 10\%$ of the X-ray sources have available {\it{Herschel}}/SPIRE photometry. 
%NOTE: The matches with HELP are 848, i.e., 848/5921=15%, but not all sources in HELP have Herschel.

As mentioned earlier, the main goal of this work is to compare the SFR of X-ray AGN with that of non-AGN systems. In particular, we focus at high X-ray luminosities ($\rm L_{X,2-10keV} > 10^{44}\,ergs^{-1}$) and examine whether the indications found by \cite{Mountrichas2021c} in the Bo$\rm \ddot{o}$tes field, are confirmed by using an X-ray sample with more than twice the number of X-ray sources at high luminosities. Lower luminosities have already been the subject of examination from  \cite{Mountrichas2022} using the COSMOS sample. AGN at low redshifts do not contribute to $\rm L_{X,2-10keV} > 10^{44}\,ergs^{-1}$. Therefore, we exclude sources below $\rm z<0.5$. Furthermore, in our SED fitting process, we use the Gaussian Aperture and Photometry (GA{\tiny {A}}P) photometry that is available in the eFEDS X-ray catalogue and the KiDS/VIKING dataset of the galaxy reference sample. GA{\tiny {A}}P photometry is performed twice, with aperture setting $\rm MIN\_APER= 0\arcsec.7$ and 1\arcsec.0. A value for each photometric band with the optimal MIN\_APER is provided \citep[for the choice of GA{\tiny {A}}P aperture size, see ][]{Kuijken2015}. GA{\tiny {A}}P is optimised for calculating {\textit{photoz}} that require colour measurements. In the case of extended and low redshift sources, total fluxes may be underestimated \citep{Kuijken2019}. For these reasons, in the following analysis we will use sources at $\rm z>0.5$.

\subsection{Galaxy reference catalogue}

To compare the SFR of X-ray AGN with non-AGN systems in a consistent manner, we compile a galaxy (non-AGN) reference catalogue. We require the same photometric coverage and apply the same SED fitting analysis in both datasets (see next section). We use the fourth data release catalogue of the KiDS/VIKING imaging survey \citep{Kuijken2019} that has available optical and NIR photometry and {\textit{photoz}} measurements for about 100 million galaxies over 1006 square degrees. We restrict the sample to those sources within the eFEDS region ($\sim 65$\,deg$^2$ overlap) and apply the same requirements for photometric coverage of the X-ray sample. We also restrict the sample to sources with $\rm z>0.5$ (see previous section). These requirements give us $\sim 200,000$ galaxies. We cross-match these sources with spectroscopic catalogues from SDSS, WiggleZ and GAMA which results in $\sim 7,000$ sources with available {\textit{specz}}. For the remaining sources we use the {\textit{photoz}} calculations provided in the KiDS/VIKING dataset. This allows us to significantly increase the size of our reference catalogue, in particular at $\rm z>1$, where a large fraction of the X-ray sources lie. {\textit{photoz}} have been estimated using the BPZ code \cite{Benitez2000}. \cite{Wright2019} tested BPZ {\textit{photoz}}, using the KiDS/VIKING photometry, against several deep spectroscopic surveys. Based on their analysis, the accuracy of {\textit{photoz}} is found at $\sigma_{\Delta z/(1+z_{spec})}=0.072$. The fraction of outliers ($\Delta z/(1+z_{{\textit{specz}}})>0.15$) is $\approx 17.7\%$. Finally, we cross-match the galaxy reference catalogue with the GAMA-09 photometric catalogue of HELP. $\sim 15\%$ of the galaxies have   been detected by {\it{Herschel}}.

\begin{table*}
\caption{The models and the values for their free parameters used by CIGALE for the SED fitting.} 
\centering
\setlength{\tabcolsep}{1.mm}
\begin{tabular}{cc}
       \hline
Parameter &  Model/values \\
	\hline
\multicolumn{2}{c}{Star formation history: delayed model and recent burst} \\
Age of the main population & 1500, 2000, 3000, 4000, 5000 Myr \\
e-folding time & 200, 500, 700, 1000, 2000, 3000, 4000, 5000 Myr \\ 
Age of the burst & 50 Myr \\
Burst stellar mass fraction & 0.0, 0.005, 0.01, 0.015, 0.02, 0.05, 0.10, 0.15, 0.18, 0.20 \\
\hline
\multicolumn{2}{c}{Simple Stellar population: Bruzual \& Charlot (2003)} \\
Initial Mass Function & Chabrier (2003)\\
Metallicity & 0.02 (Solar) \\
\hline
\multicolumn{2}{c}{Galactic dust extinction} \\
Dust attenuation law & Charlot \& Fall (2000) law   \\
V-band attenuation $A_V$ & 0.2, 0.3, 0.4, 0.5, 0.6, 0.7, 0.8, 0.9, 1, 1.5, 2, 2.5, 3, 3.5, 4 \\ 
\hline
\multicolumn{2}{c}{Galactic dust emission: Dale et al. (2014)} \\
$\alpha$ slope in $dM_{dust}\propto U^{-\alpha}dU$ & 2.0 \\
\hline
\multicolumn{2}{c}{AGN module: SKIRTOR)} \\
Torus optical depth at 9.7 microns $\tau _{9.7}$ & 3.0, 7.0 \\
Torus density radial parameter p ($\rho \propto r^{-p}e^{-q|cos(\theta)|}$) & 1.0 \\
Torus density angular parameter q ($\rho \propto r^{-p}e^{-q|cos(\theta)|}$) & 1.0 \\
Angle between the equatorial plan and edge of the torus & $40^{\circ}$ \\
Ratio of the maximum to minimum radii of the torus & 20 \\
Viewing angle  & $30^{\circ}\,\,\rm{(type\,\,1)},70^{\circ}\,\,\rm{(type\,\,2)}$ \\
AGN fraction & 0.0, 0.1, 0.2, 0.3, 0.4, 0.5, 0.6, 0.7, 0.8, 0.9, 0.99 \\
Extinction law of polar dust & SMC \\
$E(B-V)$ of polar dust & 0.0, 0.2, 0.4 \\
Temperature of polar dust (K) & 100 \\
Emissivity of polar dust & 1.6 \\
\hline
\multicolumn{2}{c}{X-ray module} \\
AGN photon index $\Gamma$ & 1.9 \\
Maximum deviation from the $\alpha _{ox}-L_{2500 \AA}$ relation & 0.2 \\
LMXB photon index & 1.56 \\
HMXB photon index & 2.0 \\
\hline
Total number of models (X-ray/reference galaxy catalogue) & 427,680,000/24,552,000 \\
\hline
\label{table_cigale}
\end{tabular}
\tablefoot{For the definition of the various parameters see Sect. \ref{sec_cigale}.}
\end{table*}

\section{Analysis}
\label{sec_analysis}

\subsection{CIGALE}
\label{sec_cigale}

To measure the (host) galaxy properties of the sources in our datasets, we apply SED fitting. For that, we use the CIGALE algorithm \citep{Boquien2019, Yang2020, Yang2022}. CIGALE allows for the inclusion of the X-ray flux in the fitting process and has the ability to account for the extinction of the UV and optical emission in the poles of AGN \citep{Yang2020, Mountrichas2021a, Mountrichas2021b, Buat2021}. 

For consistency with our previous similar studies in the Bo$\rm \ddot{o}$tes \citep{Mountrichas2021c} and the COSMOS \citep{Mountrichas2022} fields, we use the same grid used in these works. This minimises any systematic effects that may be introduced due to the different modules and parametric grid used in the SED fitting process. Table \ref{table_cigale} present the templates and the values for the free parameters. In summary, to fit the galaxy component, we utilized a delayed star formation history (SFH) model with a function form $\rm SFR\propto t \times exp(-t/\tau$). A star formation burst is included \citep{Ciesla2017, Malek2018, Buat2019} as a constant ongoing star formation of 50\,Myr. Stellar emission is modelled using the single stellar population templates of \cite{Bruzual_Charlot2003} and is attenuated following \cite{Charlot_Fall_2000}. The emission of the dust heated by stars is modelled based on \cite{Dale2014}, without any AGN contribution. The AGN emission is fit using the SKIRTOR models of \cite{Stalevski2012, Stalevski2016}. SED decomposition is able to uncover AGN that remain undetected by X-rays \citep[e.g.,][]{Pouliasis2020}. To identify such objects in the galaxy reference sample, we also include the AGN module when we fit the SEDs of these sources.

\subsection{Quality and reliability examination of the fitting results}
\label{sec_bad_fits}

To exclude from the analysis sources that are badly fitted, we impose a  reduced $\chi ^2$  threshold, $\chi ^2_r <5$. This threshold is based on visual inspection of the SEDs and has been used in previous studies \citep[e.g.,][]{Masoura2018, Mountrichas2021c, Buat2021}. $94\%$ of the X-ray AGN and $90\%$ of the galaxies in the reference catalogue satisfy this criterion. Applying a more strict criterion, e.g., $\chi ^2_r <3$, reduces the number of sources ($83\%$ of the X-ray AGN and $85\%$ of the galaxies in the reference catalogue satisfy this criterion), but does not affect the results, presented in the next section. We also exclude systems for which CIGALE could not constrain the parameters of interest (SFR, M$_*$). For that we apply the same criteria used in previous recent studies \citep[e.g.,][]{Mountrichas2021c, Mountrichas2022, Koutoulidis2021, Buat2021}. The method uses the two values that CIGALE provides for each estimated galaxy property. One value corresponds to the best model and the other value (bayes) is the likelihood weighted mean value. A large difference between the two calculations suggests a complex likelihood distribution and important uncertainties. Thus, we only include in our analysis sources with $\rm \frac{1}{5}\leq \frac{SFR_{best}}{SFR_{bayes}} \leq 5$ and $\rm \frac{1}{5}\leq \frac{M_{*, best}}{M_{*, bayes}} \leq 5$, where SFR$\rm _{best}$, M$\rm _{*, best}$ are the best fit values of SFR and M$_*$, respectively and SFR$\rm _{bayes}$ and M$\rm _{*, bayes}$ are the Bayesian values, estimated by CIGALE. $78\%$ and $88\%$ of the sources in the initial X-ray and galaxy reference catalogues meet these requirements.    

%from those sources that satisfy the M* and SFR crriteria, 94% of the X-ray and 91% of the galaxies, have x2r<5. This percentage is similar to that found for the overall population (without the SFR and M* croteria), which as I see it, shows that we need the x2r criterion to exclude badly fitted sources, regardless of the SFR and M* criteria.

In previous studies in the Bo$\rm \ddot{o}$tes \citep{Mountrichas2021c}, the XMM-XXL \citep{Mountrichas2021b} and the COSMOS \citep{Mountrichas2022} fields, it has been demonstrated that CIGALE can provide reliable SFR and M$_*$ measurements for AGN and galaxies with the same photometric coverage as in the present work, at similar redshifts. We repeat these tests for our samples and reach similar conclusions. Furthermore, the aforementioned studies have shown that lack of far-IR photometry does not affect the SFR calculation of CIGALE. Using the $10\%$ of our X-ray AGN and $15\%$ of the galaxies in the reference sample, with {\it{Herschel}} detection, we confirm these previous findings.

Finally, throughout our analysis, we take into account the uncertainties of the SFR and M$_*$ calculations, provided by CIGALE. Specifically, we calculate the significance ($\sigma=\rm value/uncertainty$) of each stellar mass, $\sigma _{\rm M_*}$, and SFR, $\sigma _{\rm SFR}$, measurement and weight each source based on these values. The total weight, $w_{\rm t}$, assigned to each source is given by the equation

\begin{equation}
w_{\rm t}= \sigma _{\rm M_*} \times \sigma _{\rm SFR}.
\end{equation}

\begin{table*}
\caption{Number of X-ray AGN and sources in the reference galaxy catalogue, after applying the mass completeness limits at each redshift interval. In the parentheses we quote the number of sources, when we also exclude quiescent systems.}
\centering
\setlength{\tabcolsep}{1.mm}
\begin{tabular}{cccc}
 \hline
&total &$\rm 0.5<z<1.0$ & $\rm 1.0<z<1.5$   \\
$\rm log (M_*/M_\odot)$ & &  $> 9.95$ & $ > 10.67$ \\
  \hline
X-ray catalogue & 1,867 (1,763) & 1,145 (1,092) & 722 (671)  \\
reference galaxy catalogue & 17,783 (17,305) & 15,261 (14,926) & 2,522 (2,379) \\
  \hline
\label{table_data}
\end{tabular}
\end{table*}

\subsection{Identification of non-X-ray AGN systems}
\label{sec_excl_agnfrac}

In the SED fitting analysis, we model the AGN emission in the case of sources in the galaxy reference catalogue, too. We use CIGALE results to identify systems with an AGN component and exclude them from the galaxy sample. Specifically, we exclude sources with $\rm frac_{AGN}>0.2$ (without taking into account the uncertainties on the $\rm frac_{AGN}$ estimates), consistently with our previous studies \citep{Mountrichas2021c, Mountrichas2022}. $\rm frac_{AGN}$ is defined as the ratio of the AGN IR emission to the total IR emission of the galaxy. This excludes $\sim 50\%$ of the sources in the galaxy reference catalogue. The percentage of sources with a significant AGN component rises from $\sim 50\%$ at $\rm 0.5<z<1.5$ to $\sim 80\%$ at $\rm z>1.5$. 

This increase of the fraction of sources with an AGN component as we move to higher redshifts, was also found in \cite{Mountrichas2021c, Mountrichas2022}, in Bo$\rm \ddot{o}$tes and COSMOS. However, in these studies the percentage was ranging from $(15-20)\%$, at $\rm 0.5<z<1.5$ to $(50-60)\%$, at $\rm z>1.5$. The higher percentages we find using the eFEDS dataset can be explained by the different M$_*$ distributions in the three samples \citep[e.g.,][]{Georgakakis2017b, Aird2018}. At $\rm 0.5<z<1.5$, galaxies from the COSMOS and Bo$\rm \ddot{o}$tes catalogues, have mean $\rm log\,[M_*(M_\odot)] = 10.5$ and 10.7, respectively. However, galaxies in the eFEDS field, at the same redshift interval are more massive, with mean $\rm log\,[M_*(M_\odot)] = 11.1$. At $\rm z>1.5$, there is also a difference of the mean M$_*$, for all three fields. Specifically, $\rm log\,[M_*(M_\odot)] = 10.7$ and 11.1 in COSMOS and Bo$\rm \ddot{o}$tes, respectively, while mean $\rm log\,[M_*(M_\odot)] = 11.4$. We split the eFEDS sources in stellar mass bins and confirm that the median value of $\rm frac_{AGN}$ increases, with increasing M$_*$. Specifically, for $\rm log\,[M_*(M_\odot)] = 9-10$, $10-11$ and $11-12$, the median $\rm frac_{AGN}$ is 0.05, 0.09 and 0.16, respectively. For the above M$_*$ bins, the fraction of sources with $\rm frac_{AGN}>0.2$ is $17\%$, $22\%$, $40\%$. 

We, also, note that $35\%$, $15\%$ and $7\%$ of the excluded sources have $\rm frac_{AGN}>0.2$ at $\rm 1,\,2\,and\,3\,\sigma$, respectively, i.e., $\rm frac_{AGN}-frac_{AGN, err}>0.2$, $\rm frac_{AGN}-2\times frac_{AGN, err}>0.2$, $\rm frac_{AGN}-3\times frac_{AGN, err}>0.2$. The corresponding fractions for the reference catalogue in Bo$\rm \ddot{o}$tes are $25\%$, $10\%$ and $5\%$ and in COSMOS are $30\%$, $13\%$ and $7\%$, respectively. Therefore, the accuracy with which CIGALE calculates the $\rm frac_{AGN}$ parameter is similar in the three fields.

Finally, we examine if our results are sensitive to the value of the $\rm frac_{AGN}$ we select. We apply a more strict criterion to identify sources with an AGN component, i.e., $\rm frac_{AGN}>0.3$ (without taking into account the uncertainties on the $\rm frac_{AGN}$ estimates). In this case, $\sim 30\%$ of the sources in the reference catalogue, are excluded. The distributions of SFR and M$_*$ of the remaining sources in the reference sample using $\rm frac_{AGN}>0.2$ and $\rm frac_{AGN}>0.3$ are identical. The median values of SFR and M$_*$ when we use $\rm frac_{AGN}>0.2$ and $\rm frac_{AGN}>0.3$ are: $\rm log\,[SFR (M_\odot yr^{-1})]=1.49$, $\rm log\,[M_*(M_\odot)]=11.18$ and $\rm log\,[SFR (M_\odot yr^{-1})]=1.48$, $\rm log\,[M_*(M_\odot)]=11.17$, respectively. Furthermore, we confirm that the choice of the value for the $\rm frac_{AGN}$ does not affect the results presented in the next section.

We conclude that an increased number of sources with (significant) AGN emission is found as we move to higher redshifts and more massive systems. This is in accordance with studies that traced the AGN activity using the distribution of the specific black hole accretion rate and found that the probability of a galaxy to host an AGN (AGN duty cycle) is higher at earlier epochs and for more massive galaxies \citep[e.g.,][]{Georgakakis2017b, Yang2017, Yang2018, Aird2018}.

This criterion significantly reduces the number of available sources in the galaxy reference catalogue, at $\rm z>1.5$. Therefore, we restrict our analysis to systems that lie at $\rm z<1.5$.

\subsection{Mass completeness}
\label{sec_mass_completeness}

Our goal is to compare the SFR of X-ray AGN with that of non-AGN galaxy systems, at different X-ray luminosities and redshifts. We also examine the role of stellar mass. This comparison could be affected by possible biases that may be introduced by the different mass completeness limits at different redshift intervals. To minimise these biases, we calculate the mass completeness at each redshift bin, following the method described in \cite{Pozzetti2010}. For these calculations, we use the galaxy reference catalogue due to its larger size. The same method has been applied in previous, similar studies \citep[e.g.,][]{Florez2020, Mountrichas2021c, Mountrichas2022}.

To estimate the mass completeness limits of our data, we, first, calculate the limiting stellar mass, M$_{*,lim}$, of each galaxy, using the following expression:

\begin{equation}
\rm log\,M_{*,lim} = log M_*+0.4(m-m_{lim}),
\end{equation}
where M$_*$ is the stellar mass of each source, measured by CIGALE, m is the AB magnitude of the source and m$_{\rm lim}$ is the AB magnitude limit of the survey. This expression estimates the mass the galaxy would have if its apparent magnitude was equal to the limiting magnitude of the survey for a specific photometric band. Then, we use the $\rm log M_{*,lim}$ of the 20\% of the faintest galaxies at each redshift bin. The minimum stellar mass at each redshift interval for which our sample is complete is the 95th percentile of $\rm log M_{*,lim}$, of the $20\%$ faintest galaxies in each redshift bin.

We use $\rm K_s$ as the limiting band of the samples, in accordance with previous studies \citep{Laigle2016, Florez2020, Mountrichas2021c, Mountrichas2022} and set $m_{lim}=21.2$ \citep{Hildebrandt2020, Salvato2021}. We find that the stellar mass completeness of our galaxy reference catalogue is $\rm log\,[M_{*,95\%lim}(M_\odot)]= 9.95$ and 10.67 at $\rm 0.5<z<1.0$ and $\rm 1.0<z<1.5$, respectively. Using the J or H NIR bands does not change significantly the mass completeness limits. Specifically, we find that $\rm log\,[M_{*,95\%lim}(M_\odot)]= 9.91$ and 9.97, at $\rm 0.5<z<1.0$, for the J and H bands, respectively and $\rm log\,[M_{*,95\%lim}(M_\odot)]= 10.74$ and 10.57, for J and H, at $\rm 1.0<z<1.5$. We confirm that using, any other near-IR band does not affect the overall results and conclusions of our work. This is also true, if we use a more dense redshift grid ($\Delta z=0.1$) to calculate the mass completeness of our dataset.

\begin{figure*}
\centering
\begin{subfigure}{0.45\textwidth}
  \centering
   \includegraphics[width=1.0\linewidth, height=6.cm]{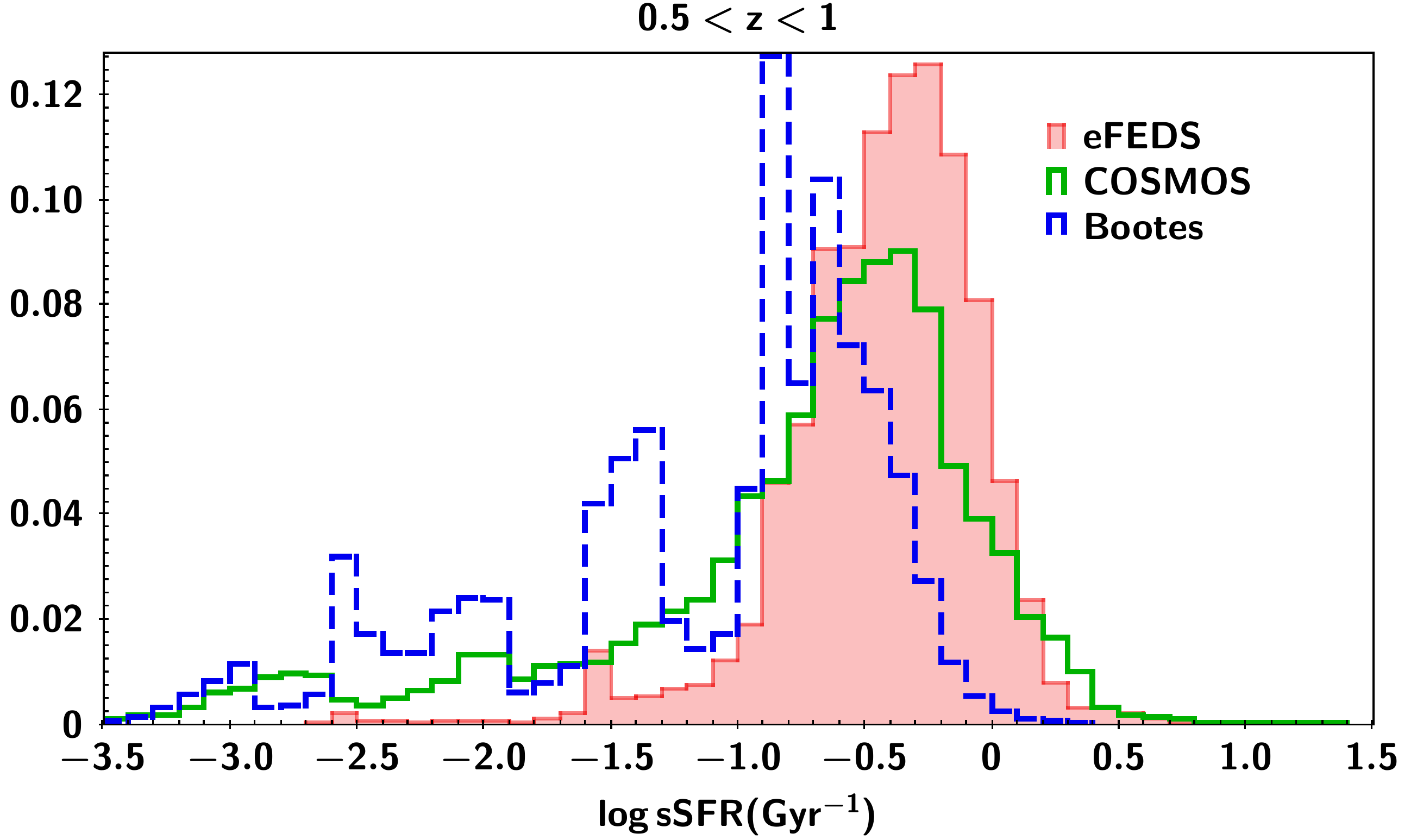}
%  \caption{}
  \label{}
\end{subfigure}
\begin{subfigure}{0.45\textwidth}
  \centering
  \includegraphics[width=1.0\linewidth, height=6.cm]{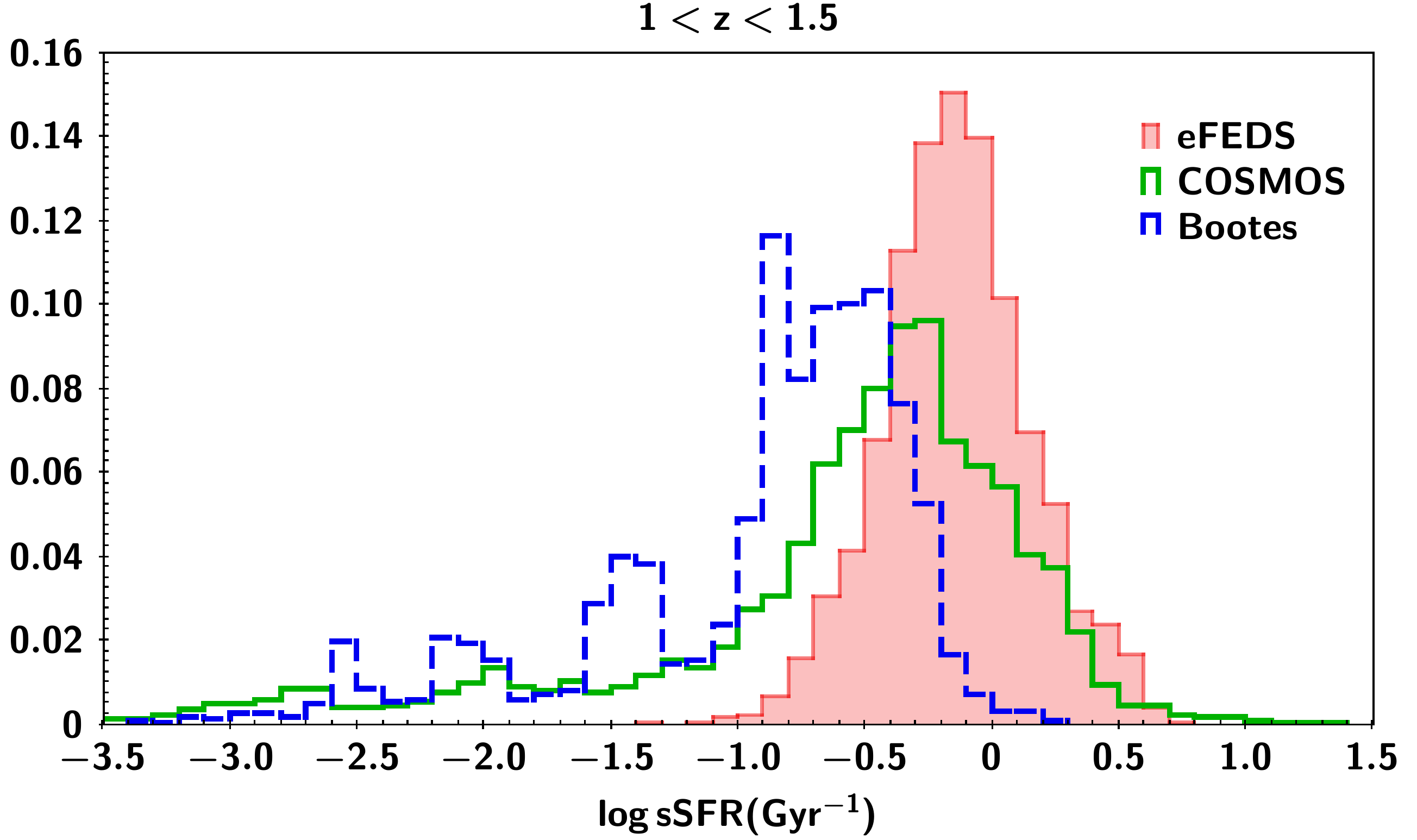}
%  \caption{}
  \label{}
\end{subfigure}
%\begin{subfigure}{0.45\textwidth}
%  \centering
%  \includegraphics[width=1.0\linewidth, height=6.cm]{AGN_sSFR_0p5_1_v3.pdf}
%%  \caption{}
%  \label{}
%\end{subfigure}
%\begin{subfigure}{0.45\textwidth}
%  \centering
%  \includegraphics[width=1.0\linewidth, height=6.cm]{AGN_sSFR_1_1p5_v3.pdf}
%%  \caption{}
%  \label{}
%\end{subfigure}
\caption{sSFR distributions, in two redshift intervals, for the galaxy reference catalogue, for sources in the eFEDS field (red shaded histogram). For comparison, we have overplotted the distributions from the COSMOS \citep[green line;][]{Mountrichas2022} and Bo$\rm \ddot{o}$tes \citep[blue line;][]{Mountrichas2021c}, fields. Sources in eFEDS do not present the long tails at small sSFR values that are observed for galaxies in COSMOS and Bo$\rm \ddot{o}$tes.}
\label{fig_ssfr}
\end{figure*}

\begin{figure}
\centering
  \includegraphics[width=\columnwidth, height=7.5cm]{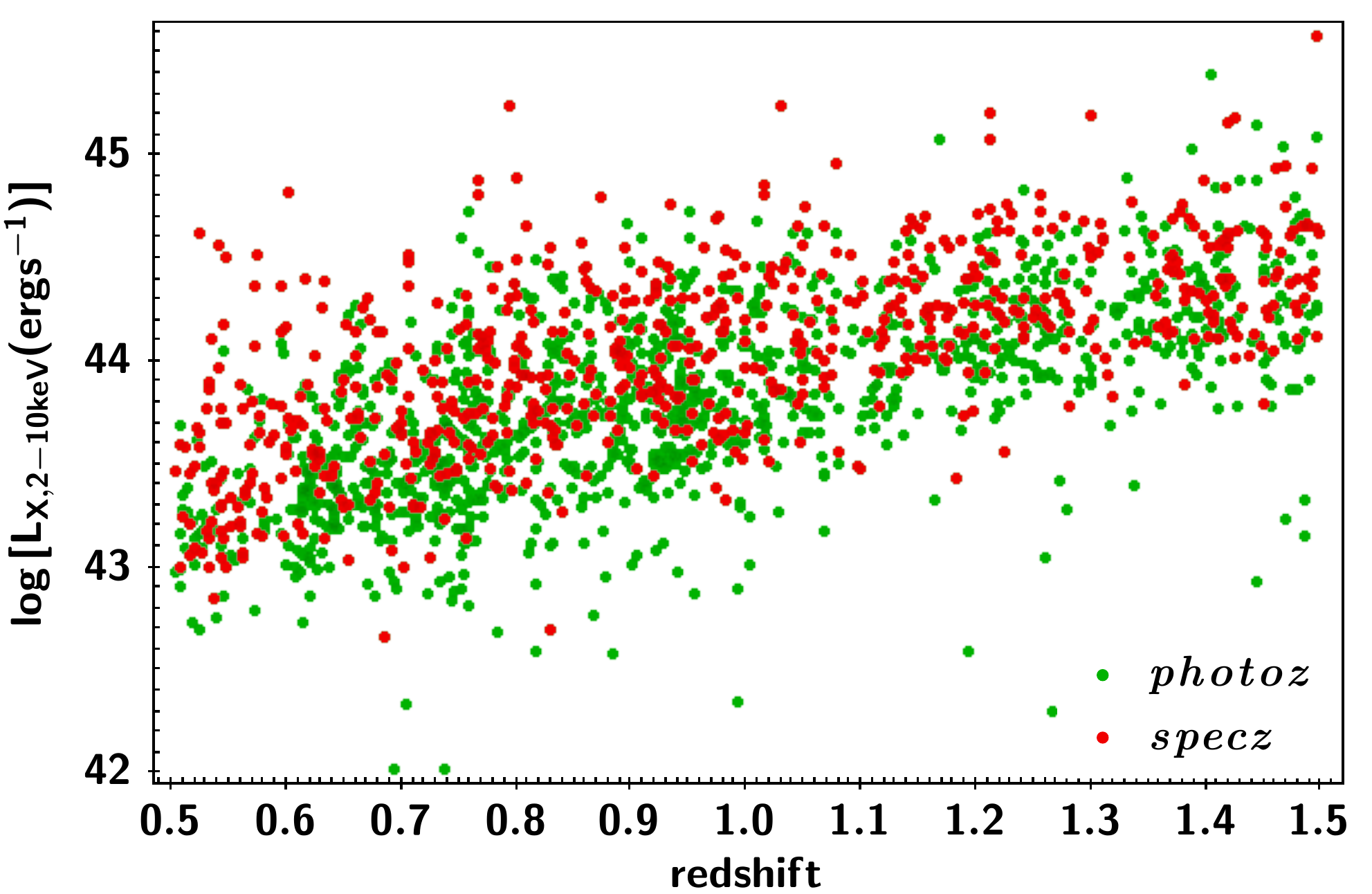} \\
  \includegraphics[width=\columnwidth, height=7.5cm]{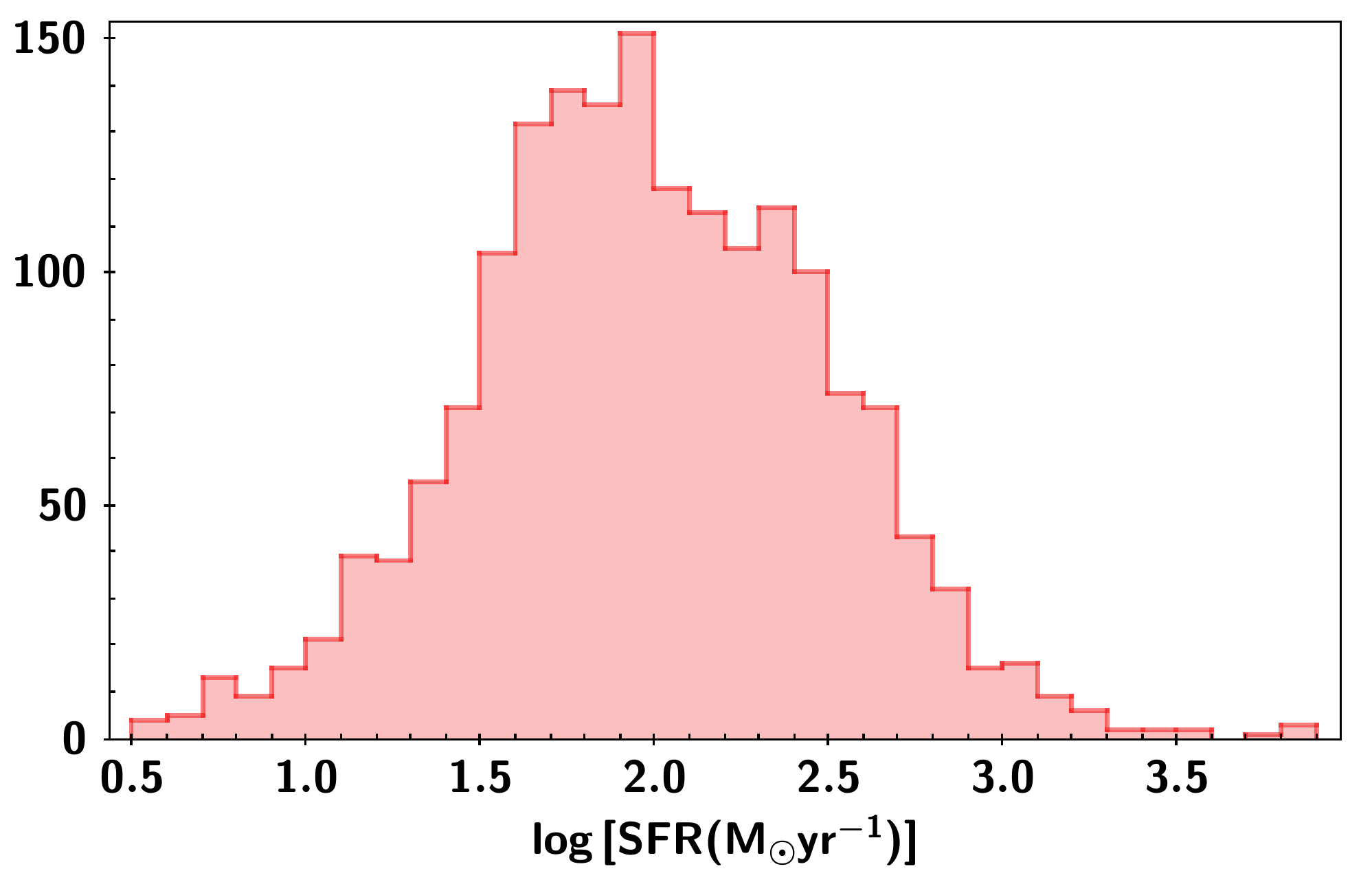} \\
  \includegraphics[width=\columnwidth, height=7.5cm]{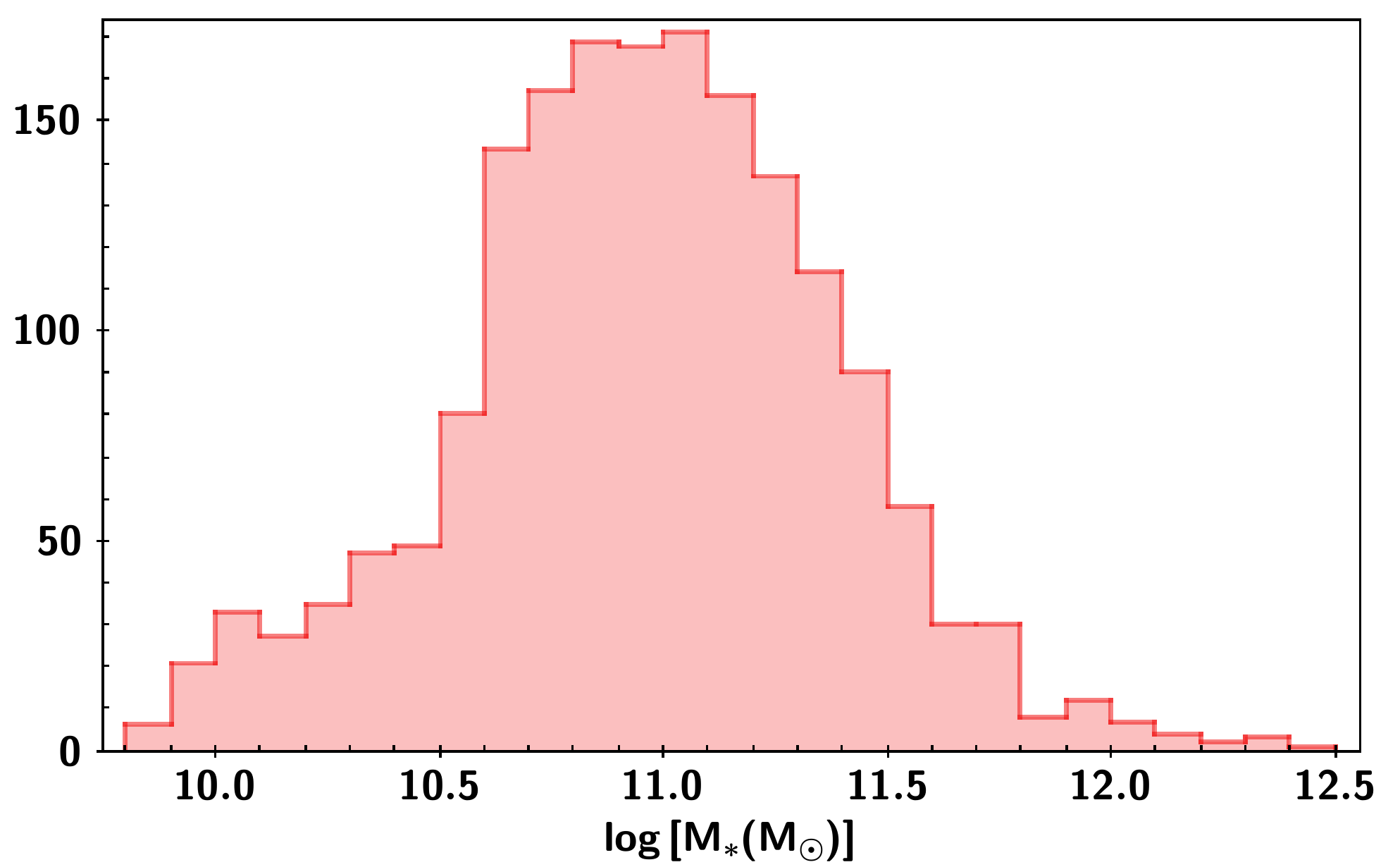}
  \caption{Top panel: The X-ray hard, intrinsic luminosity as a function of redshift, for the 1,763 X-ray AGN in our sample. 669 of them ($\sim 38\%$) have {\textit{{\textit{specz}}}}. Middle panel: SFR distribution of the 1,763 X-ray AGN. Bottom panel: M$_*$ distribution of the X-ray sources.}
  \label{fig_lx_redz}
\end{figure}

\subsection{Exclusion of quiescent systems}
\label{sec_MS}

%Previous studies \citep[e.g.,][]{Masoura2018, Bernhard2019}, compared the SFR of X-ray AGN with that of non-AGN systems, by utilizing, for the latter, the analytical formula of \cite{Schreiber2015}: 

%\begin{eqnarray}
%\log_{10}({\rm SFR}_{\rm MS} [{\rm M}_\odot / {\rm yr}]) =  m - m_0 + a_0\,r \hspace{2.5cm} \nonumber \\
%\hspace{0.9cm} - a_1 \, \big[{\rm max}(0, m - m_1 - a_2\,r)\big]^2\,,
%\label{EQ:sfrms}
%\end{eqnarray}
%with $m_0 = 0.5 \pm 0.07$, $a_0 = 1.5 \pm 0.15$, $a_1 = 0.3 \pm 0.08$, $m_1 = 0.36 \pm 0.3,$ and $a_2 = 2.5 \pm 0.6$. r and m are defined as $r \equiv \log_{10}(1+z)$ and $m \equiv \log_{10}(M_\ast / 10^{9}\,{\rm M}_\odot).$ This formula parametrizes the SFR of MS galaxies, i.e., quiescent systems have been excluded. 

In this section, we describe how we identify and reject quiescent systems from our data. \cite{Mountrichas2021c} and \cite{Mountrichas2022} used the specific SFR, sSFR ($\rm sSFR=\frac{SFR}{M_*}$), of their reference catalogues, to define quiescent galaxies. We follow their approach, which enable us to make a more consistent and fair comparison with their measurements, in the next section.

In Fig. \ref{fig_ssfr}, we plot the distributions of the sSFR of our galaxy catalogue, in two redshifts intervals (red shaded histograms). The mean $\rm log\, sSFR$ values are $\rm log\, sSFR=-0.47\,Gyr^{-1}$ and $\rm log\, sSFR=-0.18\,Gyr^{-1}$, at $\rm 0.5<z<1.0$ and $\rm 1.0<z<1.5$, respectively. This evolution of the mean sSFR with redshift is consistent with that from MS studies \citep[see Fig. 11 in][]{Schreiber2015}. \cite{Mountrichas2021c, Mountrichas2022} identified quiescent systems, based on the location of a second lower peak present in the sSFR distributions. Following their approach, we locate these secondary peaks in our distributions. At $\rm 0.5<z<1.0$, this lower peak is at $\rm log\, sSFR=-1.5\,Gyr^{-1}$. At $\rm 1.0<z<1.5$, the sSFR distribution does not present a second peak. We choose to apply a cut at $\rm log\, sSFR=-0.6\,Gyr^{-1}$, which is $\sim 2.5\times$ lower compared to the cut at the lower redshift bin, consistently with the shift of mean sSFR values for the two redshift ranges.  

%Although, the reference galaxy catalogue has been used for this exercise, due to its larger size, we notice that a lower second peak may also be identified at approximately the same sSFR values, in the X-ray AGN sample, at both redshift intervals (bottom panels). 

Table \ref{table_data} shows the number of sources remaining, after excluding quiescent systems, from both the X-ray and the galaxy samples. About $\sim  6\%$ and $\sim 4\%$ of AGN and sources in the reference catalogue, reside in quiescent systems. These percentages appear low compared to those found in \cite{Mountrichas2022}, in the COSMOS field ($\sim 25\%$ and $10\%$, for the AGN and galaxies in the reference sample, respectively) and in \cite{Mountrichas2021c}, for sources in Bo$\rm \ddot{o}$tes ($\sim 30\%$ for both datasets). For comparison, in Fig. \ref{fig_ssfr}, we overplot the sSFR distributions of the reference catalogues in COSMOS and Bo$\rm \ddot{o}$tes, for the same redshift intervals. We notice, that in both datasets there is a large tail that expands to lower sSFR values. This tail is less prominent in the case of eFEDS sources, for galaxies within $\rm 0.5<z<1$, and absent for sources in the highest redshift bin. Sources in the Bo$\rm \ddot{o}$tes field present the highest fraction of quiescent systems. This could be due to the high mass completeness limits of the Bo$\rm \ddot{o}$tes samples that biased these datasets towards systems with low sSFR values. On the other hand, the brighter luminosities spanned by our X-ray and reference galaxy catalogues compared to the datasets in the COSMOS field, may bias our samples against sources with low sSFR values (median value of $i=20.8$~mag in eFEDS, compared to 21.6 and 23.2 for the sources in Bo$\rm \ddot{o}$tes and COSMOS.

%We notice, that although at the same redshift intervals, our sSFR distributions peak at similar values with those in the COSMOS field (see Fig. 3, in Mountrichas et al. 2022), our sSFR distributions do not present the tails at low values, shown in the distributions of the COSMOS sample. The brighter luminosities spanned by our X-ray and reference galaxy catalogues compared to the datasets in the COSMOS field, may bias our samples against sources with low sSFR values (median value of $i=20.8$ in eFEDS, compared to 21.6 and 23.2 for the sources in Bo$\rm \ddot{o}$tes and COSMOS). The fraction of quiescent galaxies found by \cite{Mountrichas2021c} in the Bo$\rm \ddot{o}$tes field is even higher compared to those in this work and in the COSMOS field. However, this may be due to the high mass completeness limits of the Bo$\rm \ddot{o}$tes samples that biased these datasets towards systems with low sSFR values.

We explore other possible methods to exclude quiescent systems. Alternatively, we identify as quiescent those sources that have sSFR 1\,dex below the mean value, at each redshift range \citep[e.g.,][]{Salim2018}. Following this approach, we exclude sources with $\rm log\, sSFR<-1.4\,Gyr^{-1}$ and $\rm log\, sSFR<-1.15\,Gyr^{-1}$, at $\rm 0.5<z<1.0$ and $\rm 1.0<z<1.5$, respectively. Regarding the low redshift interval, the sSFR value is practically the same with that we used in our analysis above. At the high redshift range, the indicated sSFR value does not exclude nearly any quiescent system. 

Next, we try a more strict definition to select quiescent systems. We exclude those sources that have sSFR values 0.3\,dex below the mean value, at each redshift range. The results from following these approach are presented in Appendix \ref{appendix_quiescent}. We find that our results do not change (same trends are observed) regardless of how quiescent systems are defined. Following the more strict definition, SFR$_{norm}$ values are slightly higher, by on average only $6\%$, and well within the quoted errors of the two measurements. This effect will be discussed further in the next section.

Based on these results, in the following analysis, we exclude quiescent systems from our datasets, using the location of the secondary, low peaks of the sSFR distributions. The number of sources in our final samples are shown in Table \ref{table_data}. The top panel of Fig. \ref{fig_lx_redz} presents the position of the X-ray  AGN in the L$_X$-redshift plane. There are 751 ($\sim 50\%$ of them with {\textit{{\textit{specz}}}}) X-ray sources with $\rm L_{X,2-10keV} > 10^{44}\,ergs^{-1}$. This number is $\approx 2.2\times$ higher compared to the corresponding number in the Bo$\rm \ddot{o}$tes field \citep{Mountrichas2021c}. The middle and bottom panels of Fig. \ref{fig_lx_redz}, show the SFR and M$_*$ distributions of our X-ray sample.  The median SFR and M$_*$ values are $\rm log\,[SFR (M_\odot yr^{-1})]=1.97$ and $\rm log\,[M_*(M_\odot)]=11.07$, respectively.

\begin{figure}
\centering
  \includegraphics[width=1.\linewidth, height=8.cm]{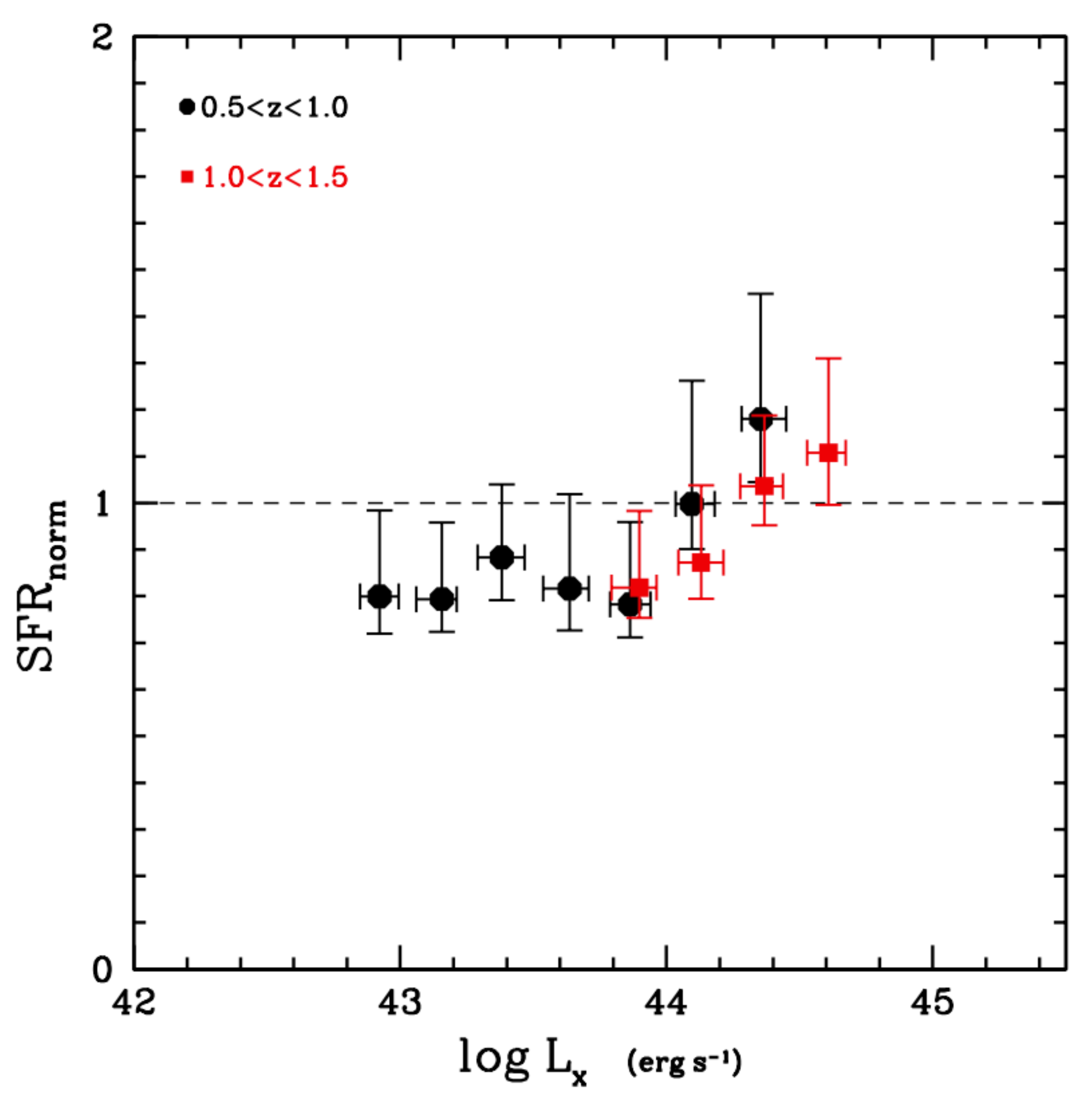}
  \caption{SFR$_{norm}$ vs. X-ray luminosity. SFR$_{norm}$ and L$_X$ are the median values of our binned measurements, in bins of L$_X$, with 0.25\,dex width. Errors are calculated using bootstrap resampling, by performing 1000 resamplings with replacement, at each bin. The dashed horizontal line indicates the SFR$_{norm}$ value ($=1$) for which the SFR of AGN is equal to the SFR of star-forming galaxies. SFR$_{norm}$ values are similar, regardless of redshift, in overlapping L$_X$. AGN host galaxies have SFR that is below or similar to the SFR of MS galaxies, at $\rm L_{X,2-10keV} < 10^{44}\,ergs^{-1}$. At higher X-ray luminosities, SFR$_{norm}$ increases, at both redshift intervals.}
  \label{fig_sfrnorm_lx_redz}
\end{figure}

\begin{figure}
\centering
  \includegraphics[width=1.\linewidth, height=8.cm]{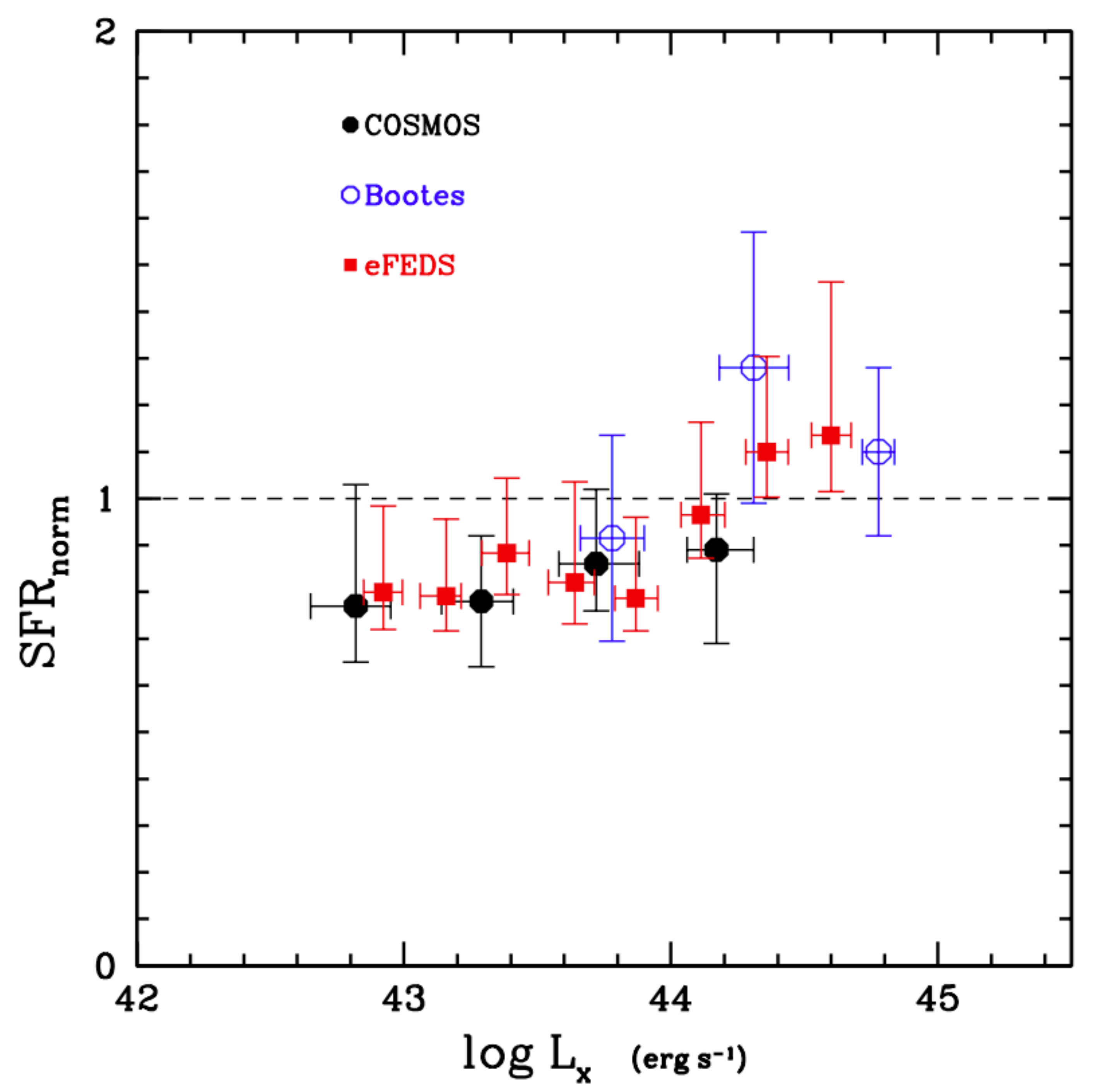}
  \caption{SFR$_{norm}$ vs. X-ray luminosity, for X-ray AGN in COSMOS, Bo$\rm \ddot{o}$tes and eFEDS. The dashed horizontal line indicates the SFR$_{norm}$ value ($=1$) for which the SFR of AGN is equal to the SFR of star-forming galaxies. Our measurements in eFEDS (red squares) agree with those from COSMOS, that AGN at low to moderate L$_X$ have lower, or at most equal, SFR with that of galaxies from the reference catalogue. At $\rm L_{X,2-10keV} > 10^{44.2}\,ergs^{-1}$, our results agree with those from Bo$\rm \ddot{o}$tes, for a small enhancement (by a factor of $\sim 15\%$) of the SFR of AGN compared to sources in the reference catalogue.}
  \label{fig_sfrnorm_lx_redz_3fields}
\end{figure} 

\begin{figure*}
\centering
\begin{subfigure}{.5\textwidth}
  \centering
  \includegraphics[width=1\linewidth]{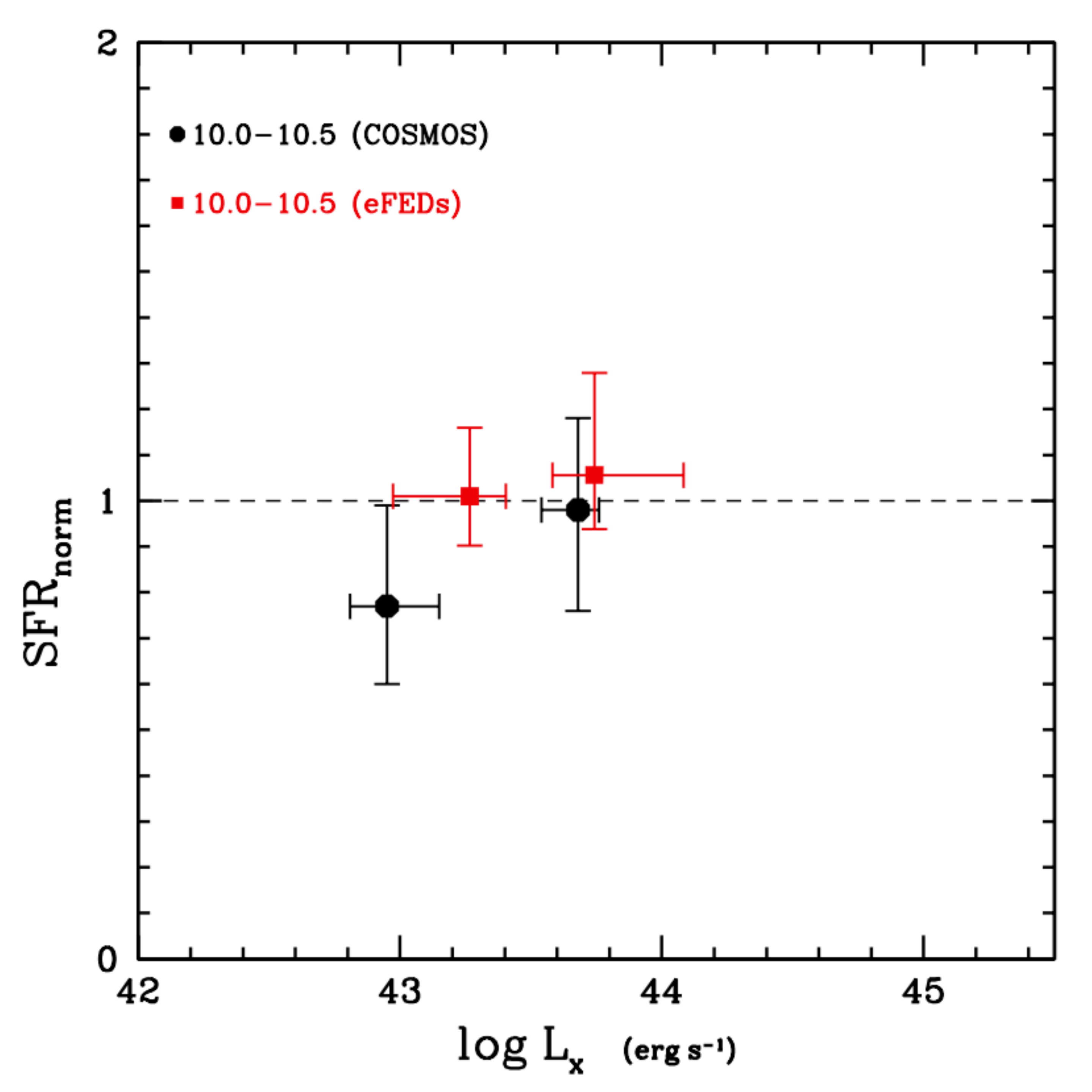}
  %\caption{A subfigure}
  %\label{fig:sub1}
\end{subfigure}%
\begin{subfigure}{.5\textwidth}
  \centering
  \includegraphics[width=1\linewidth]{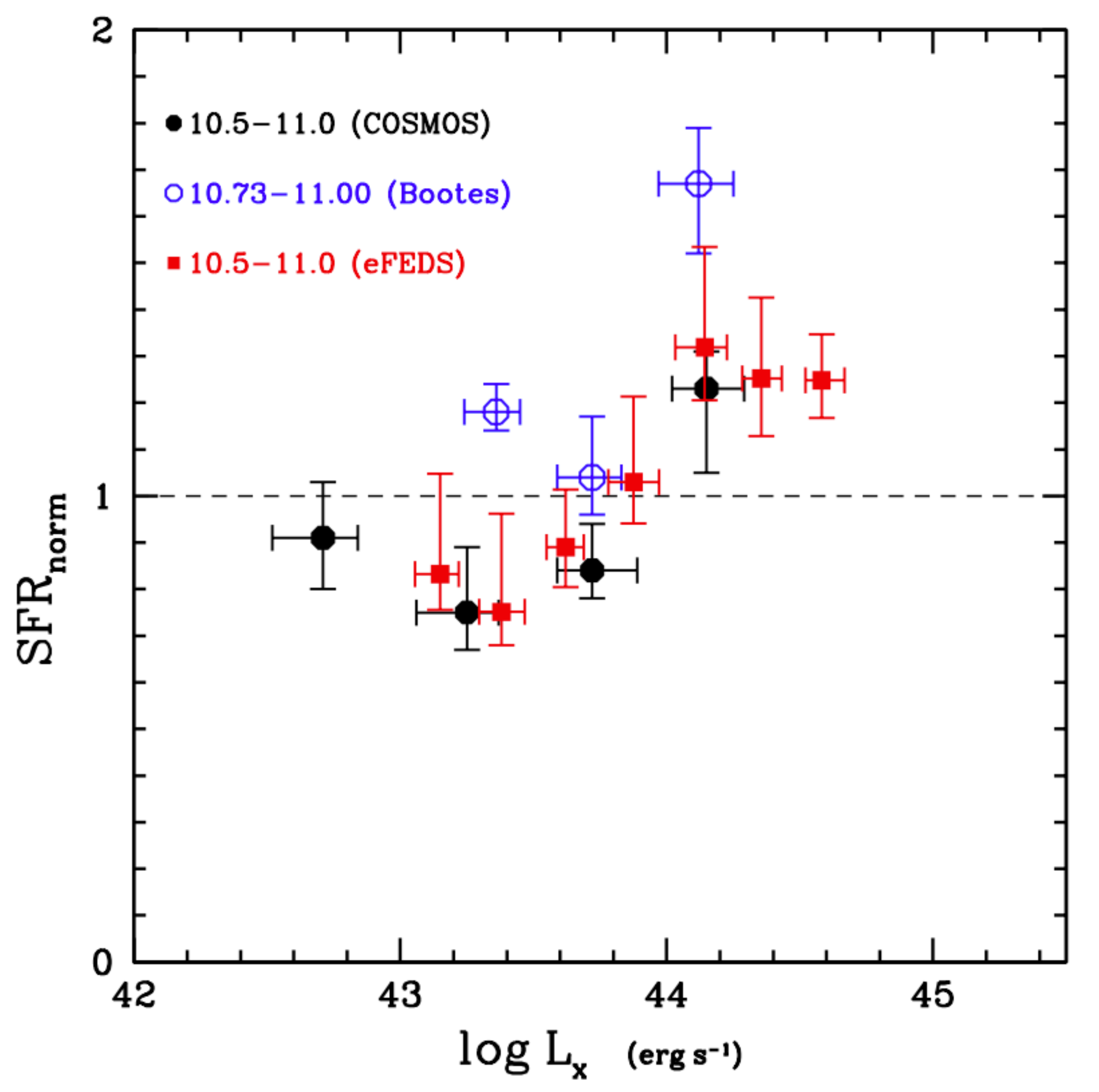}
  %\caption{A subfigure}
  %\label{fig:sub2}
\end{subfigure}
\begin{subfigure}{.5\textwidth}
  \centering
  \includegraphics[width=1.\linewidth]{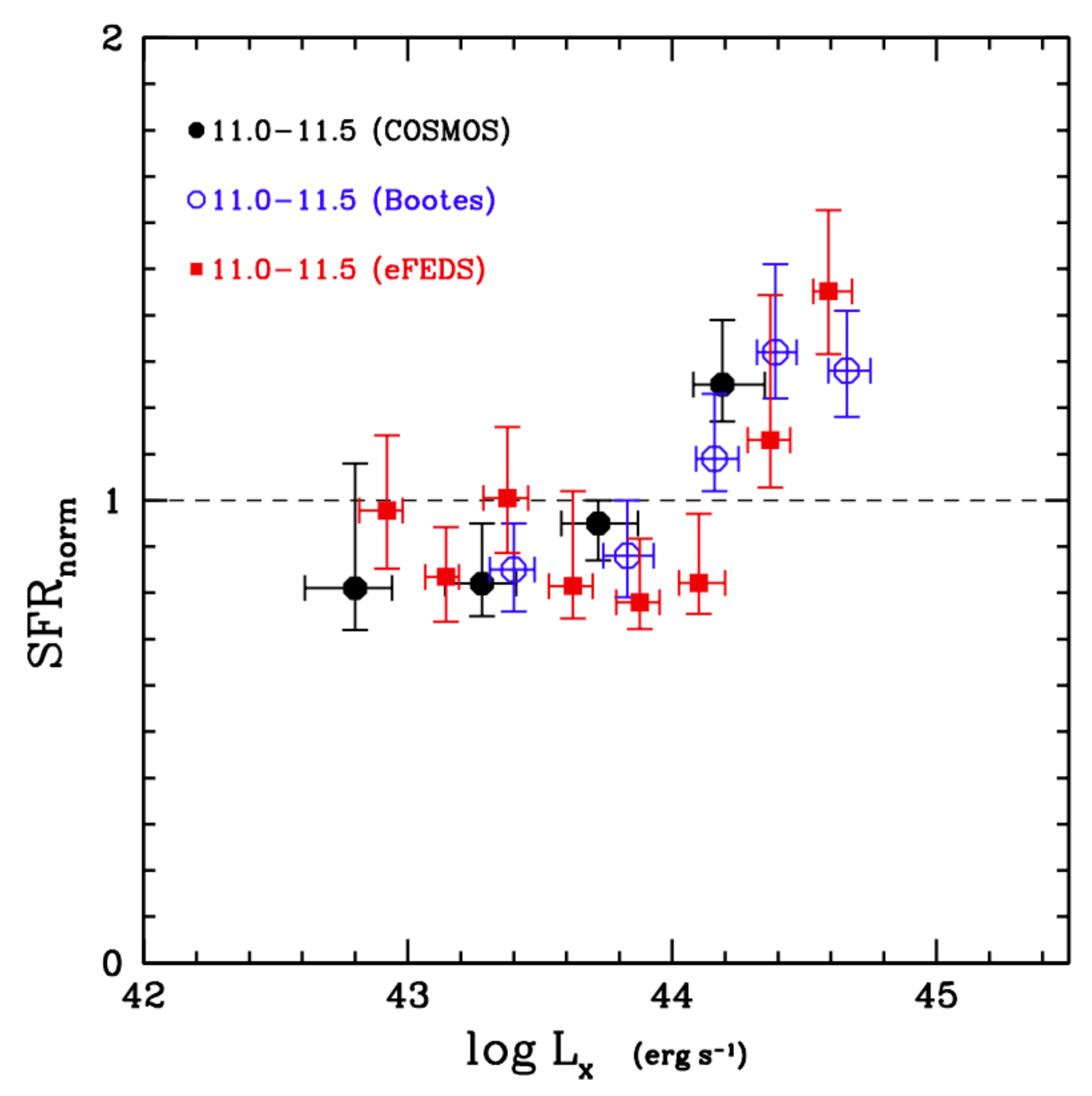}
  %\caption{A subfigure}
  %\label{fig:sub1}
\end{subfigure}%
\begin{subfigure}{.5\textwidth}
  \centering
  \includegraphics[width=1.\linewidth]{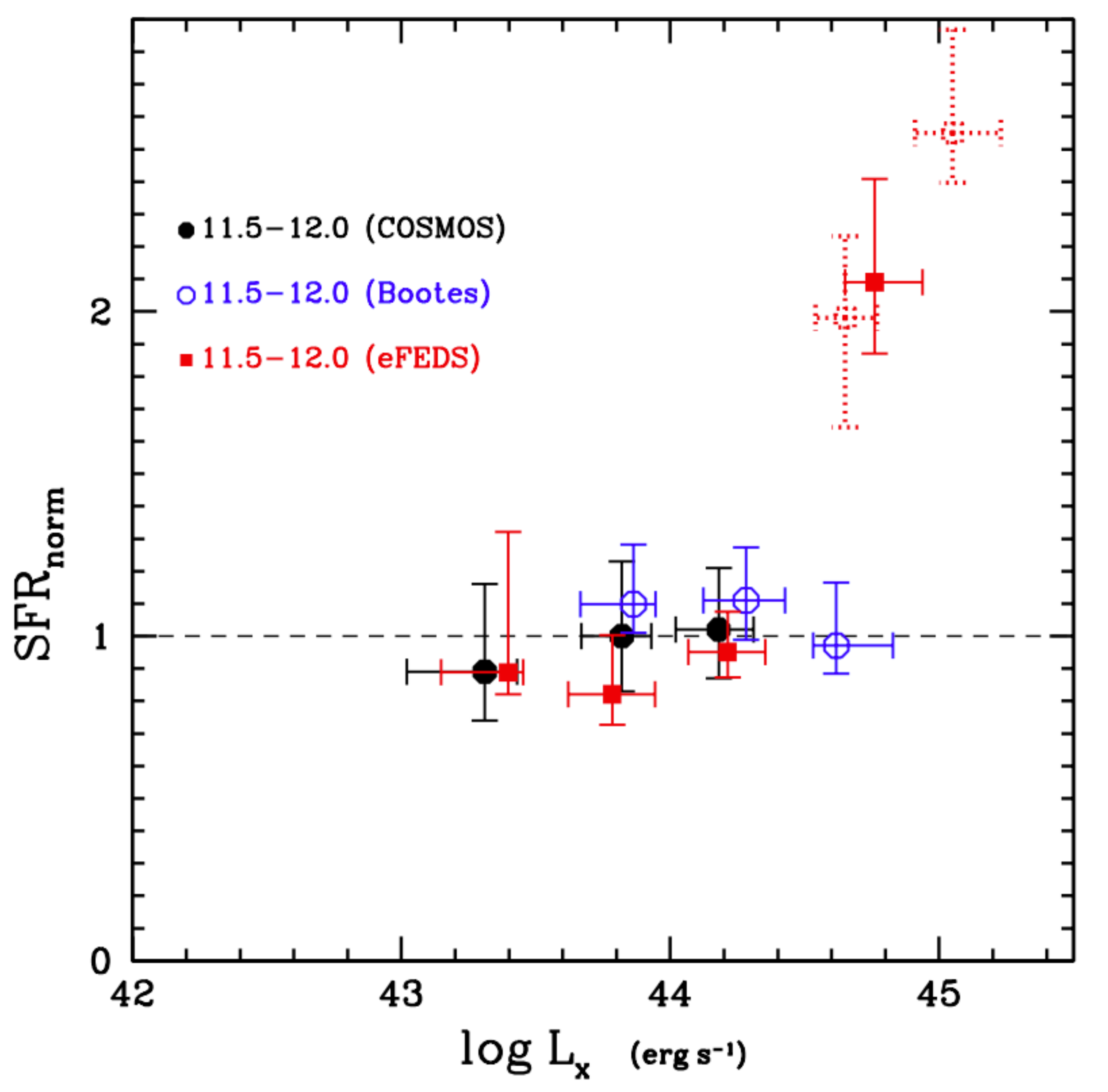}
  %\caption{A subfigure}
  %\label{fig:sub2}
\end{subfigure}
\caption{SFR$_{norm}$ vs. X-ray luminosity, for four stellar mass bins. We complement our results in the eFEDS field (red squares) with those using the COSMOS dataset \citep{Mountrichas2022} and those from the  Bo$\rm \ddot{o}$tes sample \citep{Mountrichas2021c}. The dashed horizontal line indicates the SFR$_{norm}$ value ($=1$) for which the SFR of AGN is equal to the SFR of star-forming galaxies. Errors are calculated using bootstrap resampling. In some cases the errorbars associated to the COSMOS dataset are smaller than those associated to eFEDS. This is due to the finer binning grid used for the eFEDS sample. At low luminosities ($\rm L_{X,2-10keV} < 10^{44}\,ergs^{-1}$) the results from the three studies are in very good agreement. At high luminosities ($\rm L_{X,2-10keV} > 10^{44}\,ergs^{-1}$), our measurements from the eFEDS field, confirm the tentative results from the Bo$\rm \ddot{o}$tes field, that AGN host galaxies with $\rm 10.5 < log\,[M_*(M_\odot)] < 11.5$, present enhanced SFR compared to star-forming galaxies, by $\approx 30\%$. In the most massive systems, $\rm log\,[M_*(M_\odot)] > 11.5$, our calculation show indications that, perhaps, the SFR of AGN hosts is enhanced compared to MS galaxies, only in the most luminous systems ($\rm L_{X,2-10keV} \sim 10^{45}\,ergs^{-1}$). The two dotted points show the results when we split the highest L$_X$ bin from the eFEDs field into two, using an L$_X$ cut at $\rm L_{X,2-10keV} = 10^{44.8}\,ergs^{-1}$.}
\label{fig_sfrnorm_lx_mstar}
\end{figure*}

\section{Results}
\label{sec_lx_sfr}

\subsection{Comparison of the SFR of X-ray AGN with star-forming non-AGN systems, as a function of L$_X$}

To compare the SFR of X-ray AGN with star-forming systems from the reference galaxy catalogue, we use the SFR$_{norm}$ parameter. SFR$_{norm}$ is defined as the ratio of the SFR of galaxies that host AGN to the SFR of star-forming galaxies. For its calculation, we use the SFRs that CIGALE has calculated for the 1,763 X-ray sources and the 17,305 sources in the reference galaxy. This approach has certain merits. Both samples have the same photometric coverage, the same method (SED fitting) has been applied for the estimation of (host) galaxy properties (SFR, M$_*$), the same grid has been utilized for both datasets and quiescent systems have been excluded following the same criteria for both samples. These minimize a number of systematic effects that may have affected the results of previous studies that utilized, for the estimation of the SFR of star-forming galaxies, analytical expressions from the literature \citep[e.g.,][]{Mullaney2015, Masoura2018, Bernhard2019}.

For the calculation of SFR$_{norm}$, we follow the method presented in \cite{Mountrichas2021c, Mountrichas2022}. In brief, the SFR of each X-ray AGN is divided by the SFR of sources in the reference galaxy catalogue that have M$_*$ within $\pm 0.1$ from the M$_*$ of the AGN and lie within $\rm \pm 0.075\times (1+z)$ from the AGN. Our measurements are not sensitive to the choice of the box size around the AGN. Changing the above boundaries does not change the observed trends, but affects the errors of the calculations \citep{Mountrichas2021c}. The SFR$_{norm}$ is then the median value of these ratios. In this process, each source is weighted based on the uncertainty of its SFR and M$_*$ measurement (see Sect. \ref{sec_bad_fits}). We keep only X-ray AGN for which their SFR$_{norm}$ has been measured using at least 30 sources from the reference catalogue.

In Fig. \ref{fig_sfrnorm_lx_redz}, we plot the SFR$_{norm}-$L$_X$ relation, for the two redshift intervals, we use in our analysis. In overlapping X-ray luminosities, we do not find dependence of SFR$_{norm}$ on redshift. This is in agreement with previous studies \citep{Mullaney2015, Mountrichas2021c, Mountrichas2022}. Based on this result, in Fig. \ref{fig_sfrnorm_lx_redz_3fields}, we plot SFR$_{norm}-$L$_X$ in the total redshift range spanned by our datasets, i.e., at $\rm 0.5<z<1.5$. Our goal is to examine the SFR of X-ray AGN relative to the SFR of star-forming galaxies, as a function of the AGN activity, i.e., the X-ray luminosity. We also present, in Fig. \ref{fig_sfrnorm_lx_redz_3fields}, the results from the Bo$\rm \ddot{o}$tes \citep[0.5<z<2.0;][]{Mountrichas2021c} and the COSMOS \citep[0.5<z<2.5;][]{Mountrichas2022} fields. At low to intermediate X-ray luminosities ($\rm L_{X,2-10keV} < 10^{43.5}\,ergs^{-1}$) our measurements are in very good agreement with those from COSMOS. Both results show that, in this L$_X$ regime, SFR$_{norm}$ values are systematically below the dashed line, i.e. the SFR of AGN is lower compared to that of star-forming galaxies. However, the difference is not statistically significant ($\approx 1\,\sigma$). This implies that galaxies that host X-ray AGN have lower or similar SFR compared to that of star-forming galaxies (SFR$_{norm} \leqslant 1$). At higher L$_X$ our measurements are in agreement with those in Bo$\rm \ddot{o}$tes, showing a small enhancement (on average $\sim 15\%$) which is only significant at a level of $\approx 1\,\sigma$.

%We restrict both the X-ray and the galaxy reference catalogues to those sources that have stellar population with age older than 2,800\,Myr, which includes the vast majority of the X-ray AGN host galaxies, but excludes non-AGN sources that have stellar populations that are younger than the vast majority of X-ray hosts. We also exclude systems with stellar ages older than 3,100\,Myr, which is the mean stellar age of the galaxies that host luminous X-ray sources ($\rm L_{X,2-10keV} > 10^{44.2}\,ergs^{-1}$). The results are plotted in Fig. \ref{fig_sfrnorm_lx_age}. We do not find any significant difference among the measurements. This shows that the stellar age is not the (only) reason for the enhanced SFR of X-ray AGN compared to galaxies in the reference catalogue. 

%differentiate

We perform an additional analysis to check if the SED fitting parameters influence the obtained results. In Appendix \ref{appendix_age_reliability}, we examine the reliability of CIGALE to constrain the age of the stellar population of X-ray AGN and sources in the reference catalogue. Based on our analysis, the algorithm cannot effectively calculate this parameter. When we fix the stellar age and rerun the SED fitting analysis, we find that, although the observed trends of the SFR$_{norm}$-L$_X$ relation are not affected, SFR$_{norm}$ values are increased by, on average, $17\%$, bringing SFR$_{norm}$ close to one. 

The SFR$_{norm}\leqslant 1$ found at $\rm L_{X,2-10keV} < 10^{44}\,ergs^{-1}$ may indicate that the SFR of AGN is lower than that of star-forming galaxies, i.e., the AGN reduces the star-formation of its host \citep[e.g.,][]{Zubovas2013, Appleby2020, Lacerda2020, Shen2020}. However, SFR$_{norm}$ values are statistically consistent with SFR$_{norm} \sim 1$ (within 1\,$\sigma$). Moreover, our analysis showed that the exact amplitude of SFR$_{norm}$ is susceptible to the analysis we follow, e.g., depends on the criteria we apply to exclude quiescent systems (Appendix \ref{appendix_quiescent}) and on the SFH template and parametric grid adopted (Appendix \ref{appendix_age_reliability}). We note, that in the case of the results from the COSMOS field, the SFR$_{norm}$ values are less affected by the method to exclude quiescent systems and do not change when we fix the stellar ages. However, these SFR$_{norm}$ measurements are, too, statistically consistent with one.

We conclude that, the SFR of AGN with $\rm L_{X,2-10keV} < 10^{44}\,ergs^{-1}$ is lower but consistent with that of star-forming galaxies. At higher L$_X$, we observe an increase of the SFR$_{norm}$ values, which is mild but in agreement with that found in previous studies \citep{Mountrichas2021c, Mountrichas2022}.

The trends found by our analysis are consistent with those from previous studies. \cite{Masoura2021} found a strong dependence of SFR on L$_X$, using X-ray AGN from the XMM-XXL dataset. Although our results do not present such a strong evolution of SFR$_{norm}$  with L$_X$, the overall trends are similar. Specifically, at low to moderate luminosities AGN tend to have lower or consistent SFR with non-AGN systems, while at higher L$_X$ the SFR of X-ray AGN appears enhanced compared to that of SF galaxies. We also note that in  \cite{Masoura2021}, they calculated SFR$_{norm}$ using the analysitcal expression of \cite{Schreiber2015}. As shown in \cite{Mountrichas2021c}, this approach may introduce systematics that could affect the overall results. \cite{Bernhard2019} found that the SFR$_{norm}$ distribution of AGN with $\rm L_{X,2-10keV} < 10^{43.3}\,ergs^{-1}$ is lower compared to that of MS galaxies, while more luminous X-ray sources have SFR that is consistent with that of MS SF systems, in agreement with our findings. \cite{Santini2012} used X-ray AGN from the GOODS-S, GOODS-N and XMM-COSMOS fields and compared their SFR with that of a mass-matched galaxy control sample. Based on their results the star formation of AGN is consistent with that of SF MS galaxies. Finally, \cite{Florez2020}, used X-ray AGN from the Stripe 82 field and compared their SFR with a sample of non X-ray galaxies. Their analysis showed that X-ray sources have higher SFR compared to their control galaxy sample, by a factor of $3-10$. Although our results agree with theirs regarding the enhancement of the SFR of luminous AGN compared to non-AGN systems, this enhancement  is lower based on our measurements.

\subsection{SFR$_{norm}-$L$_X$ for different M$_*$}

\cite{Mountrichas2021c} found indications that the small enhancement of SFR$_{norm}$ with L$_X$, at $\rm L_{X,2-10keV} > 10^{44}\,ergs^{-1}$, becomes (more) evident when we take into account the M$_*$ of the host galaxy. Following their analysis, we then split our measurements into four M$_*$ bins. Our goal is to use the larger size of the eFEDS sample compared to the Bo$\rm \ddot{o}$tes X-ray catalogue and add more datapoints (bins) at the high L$_X$ regime. This will allows us to see whether this increase of SFR$_{norm}$ is systematic. Our measurements also include a significantly larger number of X-ray sources in each bin which will improve the statistical significance of the measurements.

The top, left panel of Fig. \ref{fig_sfrnorm_lx_mstar}, presents SFR$_{norm}$ vs. L$_X$, for galaxies with $\rm 10.0 < log\,[M_*(M_\odot)] < 10.5$. In agreement with the results in the COSMOS  field \citep{Mountrichas2022}, the SFR of galaxies that host AGN is consistent with the SFR of star-forming galaxies (dashed line). There are no results from Bo$\rm \ddot{o}$tes in this stellar mass range, due to the mass completeness limits of the Bo$\rm \ddot{o}$tes sample. The eFEDS and COSMOS datasets do not provide us with enough AGN at  $\rm L_{X,2-10keV} > 10^{44}\,ergs^{-1}$, in this M$_*$ regime, to examine whether the SFR of X-ray AGN changes compared to star-forming galaxies, at higher L$_X$. 

In the top, right panel of Fig. \ref{fig_sfrnorm_lx_mstar}, we plot SFR$_{norm}$ as a function of L$_X$ for AGN that live in galaxies with stellar mass, $\rm 10.5 < log\,[M_*(M_\odot)] < 11.0$. We also plot the measurements from the COSMOS and Bo$\rm \ddot{o}$tes fields. For the latter, the M$_*$ interval is slightly different due to the high mass completeness values of this dataset. At low to intermediate luminosities ($\rm L_{X,2-10keV} < 10^{44.}\,ergs^{-1}$), our results are in agreement with those from previous studies and in particular with those from the COSMOS dataset, which has larger size compared to the Bo$\rm \ddot{o}$tes sample, in this luminosity interval. At $\rm L_{X,2-10keV} \approx 10^{44.2}\,ergs^{-1}$, our measurements confirm the results from the previous studies for increased SFR$_{norm}$. More importantly, the eFEDS dataset extends this trend at higher X-ray luminosities, i.e., up to $\rm L_{X,2-10keV} \approx 10^{44.6}\,ergs^{-1}$. Based on our results, X-ray AGN that live in galaxies with $\rm 10.5 < log\,[M_*(M_\odot)] < 11.0$, at $\rm L_{X,2-10keV} > 10^{44}\,ergs^{-1}$ have enhanced SFR compared to star-forming galaxies, by a $\sim 30\%$, with a statistical significance of $\approx 2\,\sigma$.

In the bottom, left panel of Fig. \ref{fig_sfrnorm_lx_mstar}, we plot SFR$_{norm}$ as a function of L$_X$ for AGN that live in galaxies with M$_*$, $\rm 11.0 < log\,[M_*(M_\odot)] < 11.5$. The trends observed are similar to those we described in the previous stellar mass bin. Specifically, at $\rm L_{X,2-10keV} > 10^{44.2}\,ergs^{-1}$, the SFR of systems that host AGN is enhanced (by $\sim 30\%$) compared to non-AGN sources. We note, that the three bins at this L$_X$ interval, from the eFEDS sample, include 326 X-ray sources compared to 128 from the Bo$\rm \ddot{o}$tes catalogue.

We conclude that X-ray AGN that live in galaxies with $\rm 10.5 < log\,[M_*(M_\odot)] < 11.5$, at $\rm L_{X,2-10keV} > 10^{44.2}\,ergs^{-1}$, present enhanced SFR (by $\sim 30\%$) compared to sources in the reference galaxy catalogue. This increase was also seen in Fig. \ref{fig_sfrnorm_lx_redz_3fields}, but when M$_*$ is taken into account the enhancement is higher and the statistical significance increases. 

%In figures \ref{fig_age} and \ref{fig_fburst}, we plot the distribution of stellar age and $f_{burst}$, for X-ray AGN with $\rm L_{X,2-10keV} > 10^{44.2}\,ergs^{-1}$ and $\rm 10.5 < log\,[M_*(M_\odot)] < 11.5$. These sub-population of AGN, presents the youngest stellar population (mean: 3012\,Myr, median: 2909\,Myr) among X-ray sources and an increased $f_{burst}$ (mean: 0.11, median: 0.12). 

The bottom, right panel of Fig. \ref{fig_sfrnorm_lx_mstar}, presents the SFR$_{norm}-$\,L$_X$ plane for galaxies with $\rm 11.5 < log\,[M_*(M_\odot)] < 12.0$. Previous studies found a flat SFR$_{norm}-$L$_X$ relation for the most massive systems. Although, our measurements are in agreement with the previous results, we find an increase of SFR$_{norm}$, SFR$_{norm}\approx 2.1$ , at $\rm L_{X,2-10keV} \approx 10^{44.8}\,ergs^{-1}$. COSMOS datapoints do not go up to such high X-ray luminosities. However, X-ray sources in Bo$\rm \ddot{o}$tes, reach similar L$_X$, but SFR$_{norm}\sim 1$. The highest L$_X$ bin from the Bo$\rm \ddot{o}$tes sample, includes 47 X-ray sources compared to 31 sources from eFEDS. Albeit, the size of the galaxy reference sample is significantly smaller in Bo$\rm \ddot{o}$tes. Specifically, in this stellar mass interval, in the Bo$\rm \ddot{o}$tes field there are 926 galaxies as opposed to 2,753 in eFEDS. As a consequence, for the calculation of the SFR$_{norm}$ of each AGN, in the Bo$\rm \ddot{o}$tes sample, each X-ray source has been matched on average with 155 galaxies compared to 495 in the case of AGN in eFEDSs. Restricting the Bo$\rm \ddot{o}$tes X-ray sample to those AGN that are matched with $>300$ galaxies from the reference sample, does not change the SFR$_{norm}$ values. The quality of the SED fits of AGN in eFEDS and in Bo$\rm \ddot{o}$tes is similar for the sources under investigation, as implied by the median $\chi ^2_{red}$ values (1.6 in eFEDS vs. 2.0 in Bo$\rm \ddot{o}$tes).

%In Fig. \ref{fig_sfrnorm_highest_bin}, we plot the distributions of the SFR$_{norm}$ values of the 51 AGN in the Bo$\rm \ddot{o}$tes field and the 31 AGN in eFEDS. The distribution of SFR$_{norm}$ for the eFEDS sources peaks at values close to the median value (SFR$_{norm} \sim 2.1$). The distribution of the Bo$\rm \ddot{o}$tes sources peaks at SFR$_{norm}\sim 0.8$, i.e. close to the median value (SFR$_{norm}\approx 0.95$). However, the distribution has a second lower peak close to the peak of the distribution of eFEDS sources. Restricting the SFR$_{norm}$ distribution of the Bo$\rm \ddot{o}$tes AGN to only those X-ray sources that are matched with $>300$ galaxies from the reference sample, does not change the shape of the distribution. 

%Fig. \ref{fig_ms_highest_bin}, presents the SFR-M$_*$ relation for the 51 and 31 X-ray AGN in the Bo$\rm \ddot{o}$tes and eFEDS fields, respectively. 

About half of the eFEDS AGN ($12/31\approx 40\%$) have $\rm log\,[SFR (M_\odot yr^{-1})]>3$ compared to $6\%$ ($3/47$) of the AGN in Bo$\rm \ddot{o}$tes. eFEDS sources also expand to $\rm L_{X,2-10keV} > 10^{45.0}\,ergs^{-1}$ (5/31), but the median L$_X$ of the two bins is similar (median $\rm L_{X,2-10keV} = 10^{44.73}\,ergs^{-1}$ and $\rm L_{X,2-10keV} = 10^{44.62}\,ergs^{-1}$, for the eFEDS and Bo$\rm \ddot{o}$tes sources, respectively) and we do not observe a correlation of SFR with L$_X$, at the narrow L$_X$ range probed by the 78 ($31+47$) AGN. Although, previous works did not find dependence of SFR$_{norm}$ with redshift \citep[e.g.,][]{Mullaney2015, Mountrichas2021c, Mountrichas2022}, we compare the redshift of the sources included in the two bins of interest. The 31 eFEDS AGN lie at $\rm 0.5<z<1.5$ (median $\rm z=1.23$), whereas the 47 X-ray sources from Bo$\rm \ddot{o}$tes are within $\rm 1.0<z<2.0$ (median $\rm z=1.67$).

Previous studies have not detected dependence of SFR$_{norm}$ with the X-ray obscuration \citep[e.g.,][]{Masoura2021, Mountrichas2021c}. Nevertheless, we examine the X-ray obscuration of the AGN as a possible source of the different results found for this particular high L$_X$ bin. 32/47 ($68\%$) of the X-ray sources in Bo$\rm \ddot{o}$tes are X-ray obscured (N$_H>21.5$\,cm$^{-2}$) compared to 10/31 ($32\%$) of the eFEDS AGN. However, we do not find a tendency for lower SFR$_{norm}$ values for the obscured sources, in any of the two fields. This is also true, if we increase the N$_H$ value used for the X-ray classification (N$_H>22$\,cm$^{-2}$). Similar results are found when we classify sources based on their inclination angle, $i$, estimated by CIGALE.  

Investigating further the properties of the sources in the two bins (e.g. AGN fraction, dust attenuation), we find similar distributions and median/average values. Their photometric coverage is also similar. eFEDS AGN, though, are more optically luminous compared to their Bo$\rm \ddot{o}$tes counterparts (median $i=19.3$ compared to 21.5).

%Finally, we compare the values of the $f_{burst}$ parameter for the 51 Xray sources from the Bo$\rm \ddot{o}$tes sample and the 31  AGN from eFEDS. The latter have significantly larger $f_{burst}$. Specifically, the median value of $f_{burst}$ for the 51 AGN in Bo$\rm \ddot{o}$tes AGN $f_{burst}=0.01$ compared to $f_{burst}=0.08$ of the 31 X-ray sources in eFEDS. These trends are independent of the AGN classification, based on the inclination angle.

We also split the highest L$_X$ bin from eFEDS into two, using a luminosity cut at $\rm L_{X,2-10keV} = 10^{44.8}\,ergs^{-1}$. 23/31 AGN have lower L$_X$ than this cut and 8/31 have higher L$_X$. The results are shown by the dotted, red squares in the bottom, right panel of Fig. \ref{fig_sfrnorm_lx_mstar}. Although the number of sources included in the two bins is small and no strong conclusions can be drawn, we observe an increase of SFR$_{norm}$ within the L$_X$ range probed by the 31 AGN. However, the lowest L$_X$ bin, of the two newly created, is still significantly higher compared to the highest L$_X$ bin from Bo$\rm \ddot{o}$tes, although the L$_X$ ranges spanned by the two bins are very similar.

A strong conclusion cannot be drawn, but we cannot rule out the possibility, the SFR of the most massive AGN hosts to be enhanced compared to MS galaxies at high L$_X$, in accordance with our findings for less massive systems, but this enhancement to occur at even higher luminosities ($\rm L_{X,2-10keV} \sim 10^{45}\,ergs^{-1}$). The reason that this trend was not observed in the Bo$\rm \ddot{o}$tes field could be a selection effect related to the significantly smaller size (by $\approx 12\times$) of that field compared to eFEDS.

\section{Summary-Conclusions}
\label{sec_summary}

We used $\sim 1,800$ X-ray selected AGN from the eFEDS field and more than $17,000$ galaxies in the same spatial volume ($\rm 0.5<z<1.5$) and compared the SFR of the two populations. Both samples have the same photometric coverage. We performed SED fitting, using the CIGALE algorithm and the same templates and parametric grid for both datasets. We accounted for the mass completeness of the two catalogues and applied a uniform method to exclude quiescent sources from both samples. These allowed us to compare the SFR of X-ray AGN and non-AGN systems in a uniform manner, minimizing systematic effects. Furthermore, our analysis and SED fitting grid are identical to those applied in previous studies \citep{Mountrichas2021c, Mountrichas2022}, that used X-ray sources from different fields (COSMOS, Bo$\rm \ddot{o}$tes) and spanned different X-ray luminosities. This allows us to compare and complement our results with theirs and draw a picture regarding the location of X-ray AGN relative to the MS, in over 2.5 order of magnitude in L$_X$ ($\rm L_{X,2-10keV} \sim 10^{42.5-45}\,ergs^{-1}$).

Our results showed that at low to moderate X-ray luminosities, $\rm L_{X,2-10keV} < 10^{44}\,ergs^{-1}$, X-ray AGN have SFR that is below, or at most equal, to that of star-forming galaxies. This is in agreement with the results of \cite{Mountrichas2022} that used X-ray sources in the COSMOS field. \cite{Mountrichas2021c} used X-ray data from the Bo$\rm \ddot{o}$tes field, that span high L$_X$, and found indications that the SFR of AGN is higher compared to that of MS galaxies, at $\rm L_{X,2-10keV} > 2-3\times 10^{44}\,ergs^{-1}$. The eFEDS sample is $2-3$ times larger at these luminosities compared to the Bo$\rm \ddot{o}$tes dataset, enabling us to increase the number of datapoints in this L$_X$ regime. Our results confirm these previous tentative results. Specifically, luminous AGN that live in galaxies with $\rm 10.5 < log\,[M_*(M_\odot)] < 11.5$ have SFR that is by $\sim 30\%$ higher than that of non-AGN star-forming galaxies.

%Based on our analysis, the stellar populations of galaxies in the reference catalogue and those that host X-ray AGN, have similar ages. However, higher luminosity AGN ($\rm L_{X,2-10keV} > 10^{44.2}\,ergs^{-1}$) appear to have enhanced SFR. compared to their lower L$_X$ counterparts and MS galaxies. One possible scenario is that in systems that host X-ray luminous AGN, the enhanced SFR is due to a recent star-formation burst that substantially increased the number of total stars of the galaxy. Investigating this hypothesis further, we found that $f_{burst}$, i.e., the fraction of stars formed in a recent burst relative to the total mass stars ever created, is higher by an order of magnitude in galaxies that host luminous X-ray AGN compared to non-AGN systems (and less luminous AGN). These trends are more evident in luminous AGN with  $\rm 10.5 < log\,[M_*(M_\odot)] < 11.5$.

Finally, for the most massive systems ($\rm 11.5 < log\,[M_*(M_\odot)] < 12.0$, we find a flat SFR$_{norm}-$L$_X$ relation up to $\rm L_{X,2-10keV} \sim 10^{44.5}\,ergs^{-1}$, with SFR$_{norm}\sim 1$. Although, this picture holds at even higher L$_X$, based on X-ray AGN from the Bo$\rm \ddot{o}$tes field \citep{Mountrichas2021c}, in our analysis, we detect significant enhancement of SFR$_{norm}$, at $\rm L_{X,2-10keV} > 10^{44.5}\,ergs^{-1}$, by a factor of $\sim 2$. Based on our investigations, we cannot rule out the possibility that in the case of the most massive AGN host galaxies, the SFR is enhanced compared to star-forming galaxies at high L$_X$, in agreement with our results for less massive systems, but this enhancement occurs at even higher L$_X$.

X-ray AGN, due to their different triggering mechanisms, constitute a diverse extragalactic population, hosted by a variety of galaxies. Our current analysis, complemented by the results from \citep{Mountrichas2021c} and \citep{Mountrichas2022}, in the Bo$\rm \ddot{o}$tes and COSMOS fields, showed that is not only important to compare the SFR of AGN host galaxies with non-AGN systems in a uniform manner, but to study it in a wide range of X-ray luminosities and galaxy properties. 

%Our previous work \citep{Mountrichas2021c}, demonstrated the importance of comparing the SFR of X-ray AGN with that of non-AGN systems in a uniform manner. It also highlighted the caveats when we compare results from different studies that have applied different photometric criteria to select their sourcees and different methods and parametric grids to measure (host) galaxy properties. X-ray AGN, due to their different triggering mechanisms, constitute a diverse extragalactic population, hosted by a variety of galaxies. Our current analysis, complemented by the results from \citep{Mountrichas2021c} and \citep{Mountrichas2022}, in the COSMOS and Bo$\rm \ddot{o}$tes fields, showed that is not only important to compare the SFR of AGN host galaxies with non-AGN systems in a uniform manner, but to study it in a wide range of X-ray luminosities and for different host galaxy properties 

\begin{acknowledgements}
GM acknowledges support by the Agencia Estatal de Investigación, Unidad de Excelencia María de Maeztu, ref. MDM-2017-0765.
The project has received funding from Excellence Initiative of Aix-Marseille University - AMIDEX, a French 'Investissements d'Avenir' programme.
MB gratefully acknowledges support by the ANID BASAL project FB210003.
K.M. is supported by the Polish National Science Centre grant UMO-2018/30/E/ST9/00082.
\end{acknowledgements}

\bibliography{mybib}{}
\bibliographystyle{aa}

\appendix

\section{Identification of quiescent systems}
\label{appendix_quiescent}

In this section, we examine whether our measurements are affected by the method we apply to exclude quiescent systems from the X-ray and galaxy reference catalogues. In our analysis, we exclude such sources based on the sSFR distributions. Specifically, we follow the approach of \cite{Mountrichas2021c, Mountrichas2022} and select quiescent systems based on the location of a second lower peak in the sSFR distributions. However, this peak is not prominent in our sample, for sources within $\rm 1.0<z<1.5$, and the fraction of quiescent galaxies identified is small. 

We apply a more strict criterion to exclude quiescent galaxies and examine its effect on our results. Specifically, we reject from our analysis sources that have sSFR that is 0.3\,dex below the mean values of the sSFR of galaxies in the reference catalogue. At $\rm 0.5<z<1.0$, we exclude systems with $\rm log\, sSFR<-0.7\,Gyr^{-1}$, while at $\rm 1.0<z<1.5$, we exclude sources with  $\rm log\, sSFR<-0.45\,Gyr^{-1}$. These criteria, identify $20\%$ of the X-ray sources and $18\%$ of the sources in the reference catalogue, as quiescent systems. 

\begin{figure}
\centering
  \includegraphics[width=1.\linewidth, height=8.cm]{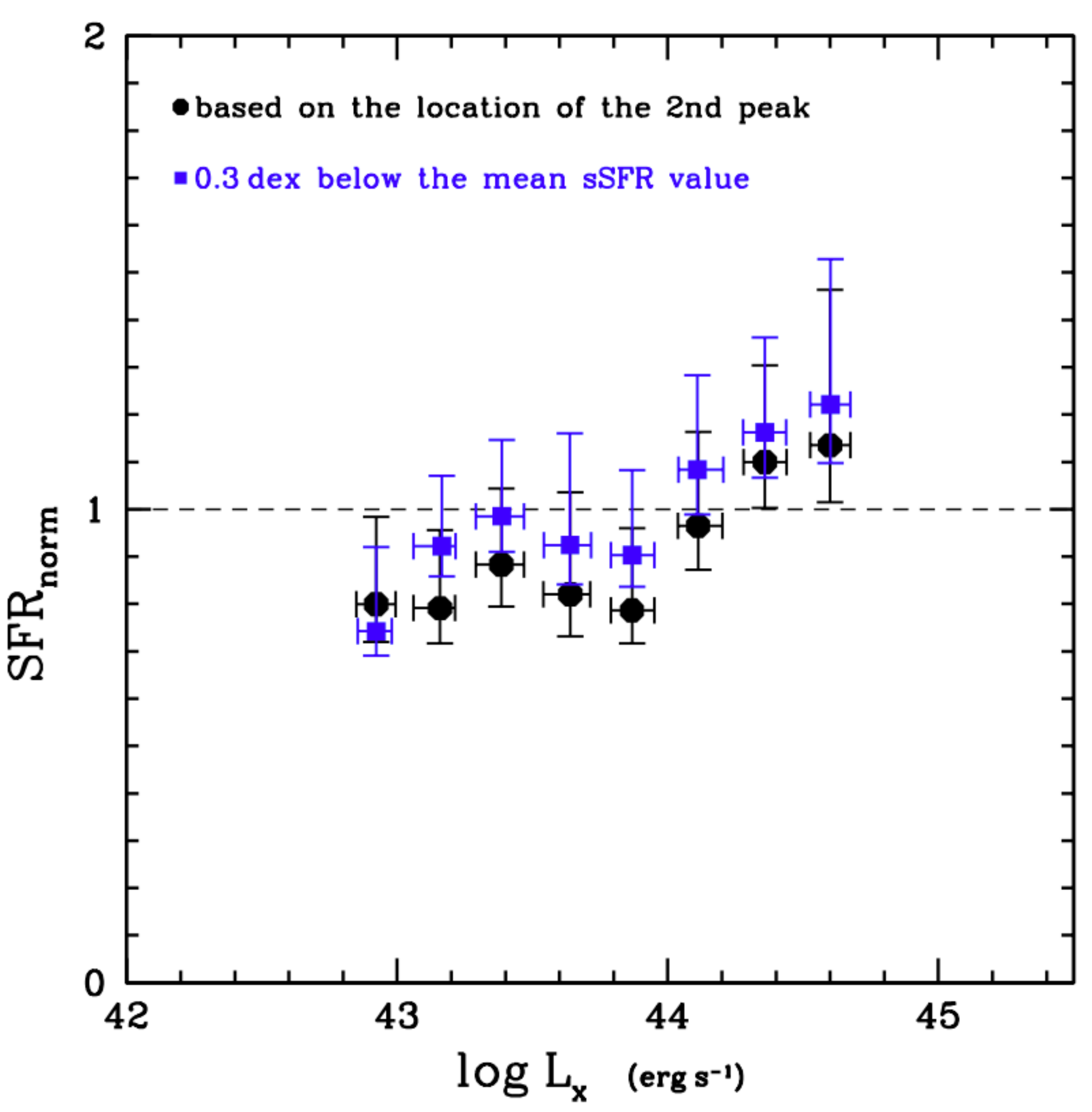}
  \caption{Comparison of the SFR$_{norm}-$L$_X$ for different definitions of quiescent systems. Black circles show the SFR$_{norm}-$L$_X$ measurements when we exclude from the X-ray sample and the galaxy reference catalogue, quiescent sources, based on the location of the lower, second peak in the sSFR distributions. Blue circles present the results when exclude as quiescent these systems that have sSFR values 0.3\,dex below the mean sSFR value. Errors are estimated via bootstrap resampling. The dashed horizontal line indicates the SFR$_{norm}$ value ($=1$) for which the SFR of AGN is equal to the SFR of star-forming galaxies.}
  \label{fig_sfrnorm_lx_quiesc}
\end{figure}

We, then, measure the SFR$_{norn}$ of each X-ray AGN, as described in Sect. \ref{sec_lx_sfr} and  bin the results in L$_X$ bins with width 0.25\,dex, for the total redshift range spanned by our datasets. The results are presented  in Fig. \ref{fig_sfrnorm_lx_quiesc}. Errors have been estimated, using bootstrap resampling. For comparison, we also plot the measurements from the samples used in our analysis, i.e., excluding quiescent systems based on the location of the lower, second peak of the sSFR distributions. We notice that the results using the more strict criterion to identify quiescent galaxies are slightly higher (by on average $\sim 6\%$). The difference, though, is marginal, i.e., the results are consistent within the errors of the two measurements. Most importantly, the observed trends are identical between the two results. In detail, SFR$_{norm}$ is below one at $\rm L_{X,2-10keV} < 10^{44}\,ergs^{-1}$ and there is a small increase at higher luminosities, where SFR$_{norm}$ becomes larger than one. This result shows that the (mild) increase of SFR$_{norm}$ we observe at high L$_X$ is not sensitive to how effectively we remove quiescent systems from our samples. 

We conclude, that the method we choose to identify quiescent systems from our X-ray and galaxy reference catalogues does not affect our overall results and conclusions.

\section{The effect of the adopted SFH on SFR$_{norm}$ calculations}
\label{appendix_age_reliability}

In this section, we examine the effectiveness of CIGALE to constrain the age of the stellar populations of AGN and sources in the reference sample and its effect on the SFR$_{norm}$ measurements. For that purpose, we use the ability of CIGALE to create mock catalogues. These catalogues can be used to assess the validity of a parameter estimation. To create them the algorithm considers the best fit of each source in the dataset. The code uses the best model flux of each galaxy and inserts a noise, extracted from a Gaussian distribution with the same standard deviation as the observed flux. Then the mock data are analysed following the same process as for the data \citep{Boquien2019}.

\begin{figure}
\centering
  \includegraphics[width=\columnwidth, height=7.cm]{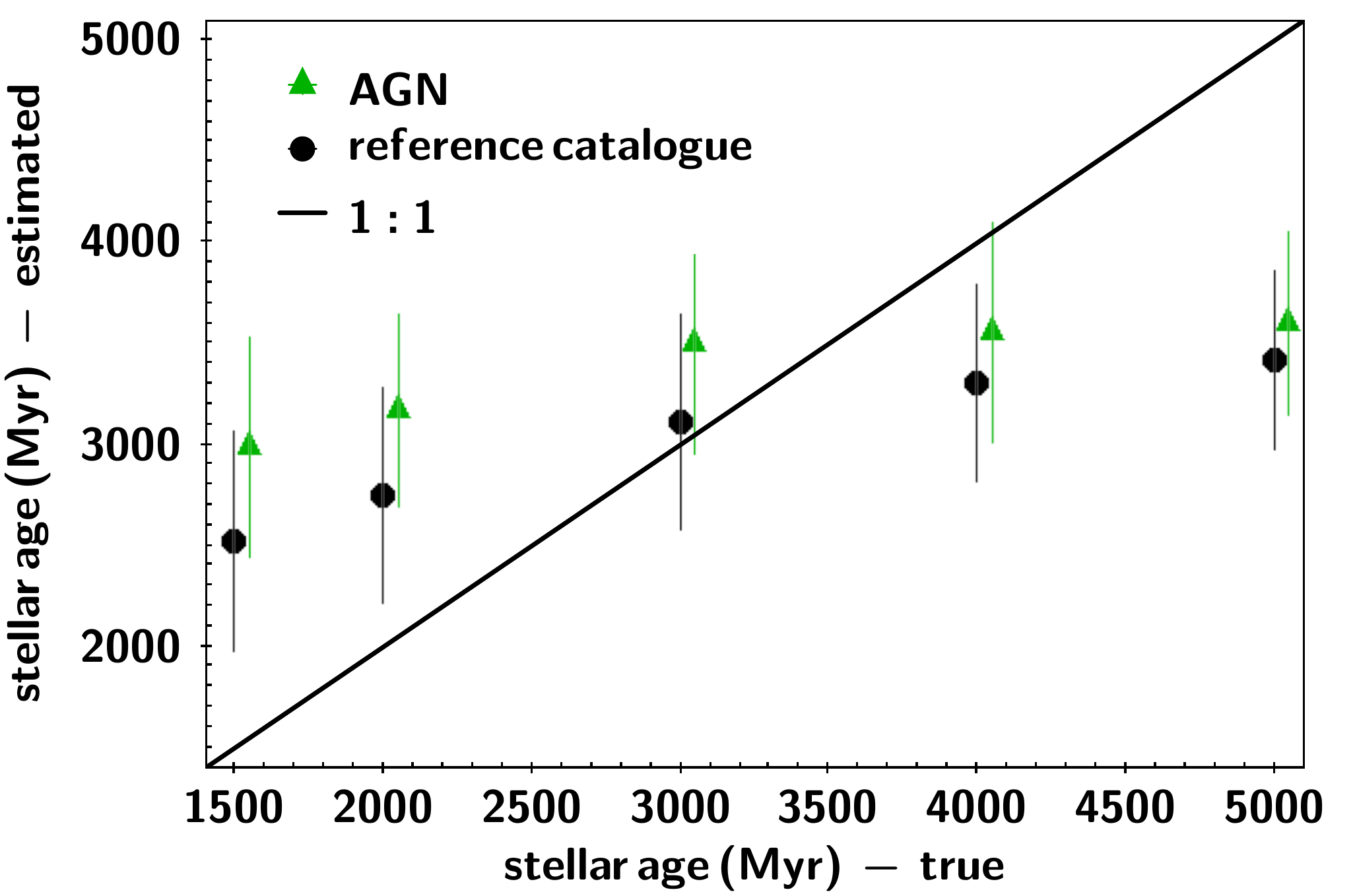}
  \caption{Comparison of the age of the main stellar population measurements of galaxies in the reference catalogue and AGN, for the estimated and true values, from the mock analysis. For both populations, CIGALE overestimates the parameter, at $<3000$\,Myr. The trend is more evident in the case of X-ray sources. The algorithm  underestimates the parameter for older stellar ages. Errors represent the standard deviation of the measurements. AGN measurements have been offset by 50\,Myr on the horizontal axis, for clarity.}
  \label{fig_age_main_mock}
\end{figure}

Based on the results, presented in Fig. \ref{fig_age_main_mock}, CIGALE cannot constrain this parameter. Specifically, the algorithm overestimates the ages of the stellar populations both for the X-ray AGN and the sources in the reference catalogue for stellar ages $\leq 3000$\,Myrs and underestimates them for ages $\geq 4000$\,Myrs. Furthermore, the parameter measurements are systematically lower in the case of non-AGN systems, at least for stellar ages $<3500$\,Myrs. We, also, examine whether these results are susceptible to the selection of the SFH module. For that purpose, we run again CIGALE, both for the X-ray and the galaxy reference catalogue, using a delayed SFH template that allows both an instantaneous recent variation of the SFR upwards (burst) and downwards \citep[quenching;][]{Ciesla2017, Boquien2019}. The results are not affected by the different SFH module.

%\section{The effect of the limited constrain of stellar age on SFR$_{norm}$}
%\label{appendix_age_fixed}

We conclude that, CIGALE cannot effectively constrain the stellar age of the sources. Now, we examine, if and how this affects the SFR$_{norm}$ calculations. For that purpose, we fix the stellar age of each system to a value that is equal to the age of the universe (in Myrs) at the redshift of the source, minus 10\%-15\% and rerun CIGALE, both for the AGN and the reference catalogues. This is based on the expectation that the first galaxies started forming stars a few hundred Myr after the big bang \citep[e.g.,][]{Oesch2016}. 

%In Figures \ref{fig_fixed_agn} and \ref{fig_fixed_reference}, we compare the SFR (top panels) and M$_*$ (bottom panels) measurements, for the X-ray AGN and sources in the reference catalogue, respectively, when the stellar age is free and fixed, during the SED fitting. Similar trends are observed for both datasets, although, in the case of galaxies in the reference catalogue these trends are more evident. Specically, SFR measurements tend to be higher and M$_*$ lower when the stellar age is set free. We, then, examine how this affects the SFR$_{norm}$ calculations. 

\begin{figure}
\centering
  \includegraphics[width=1.\linewidth, height=8.cm]{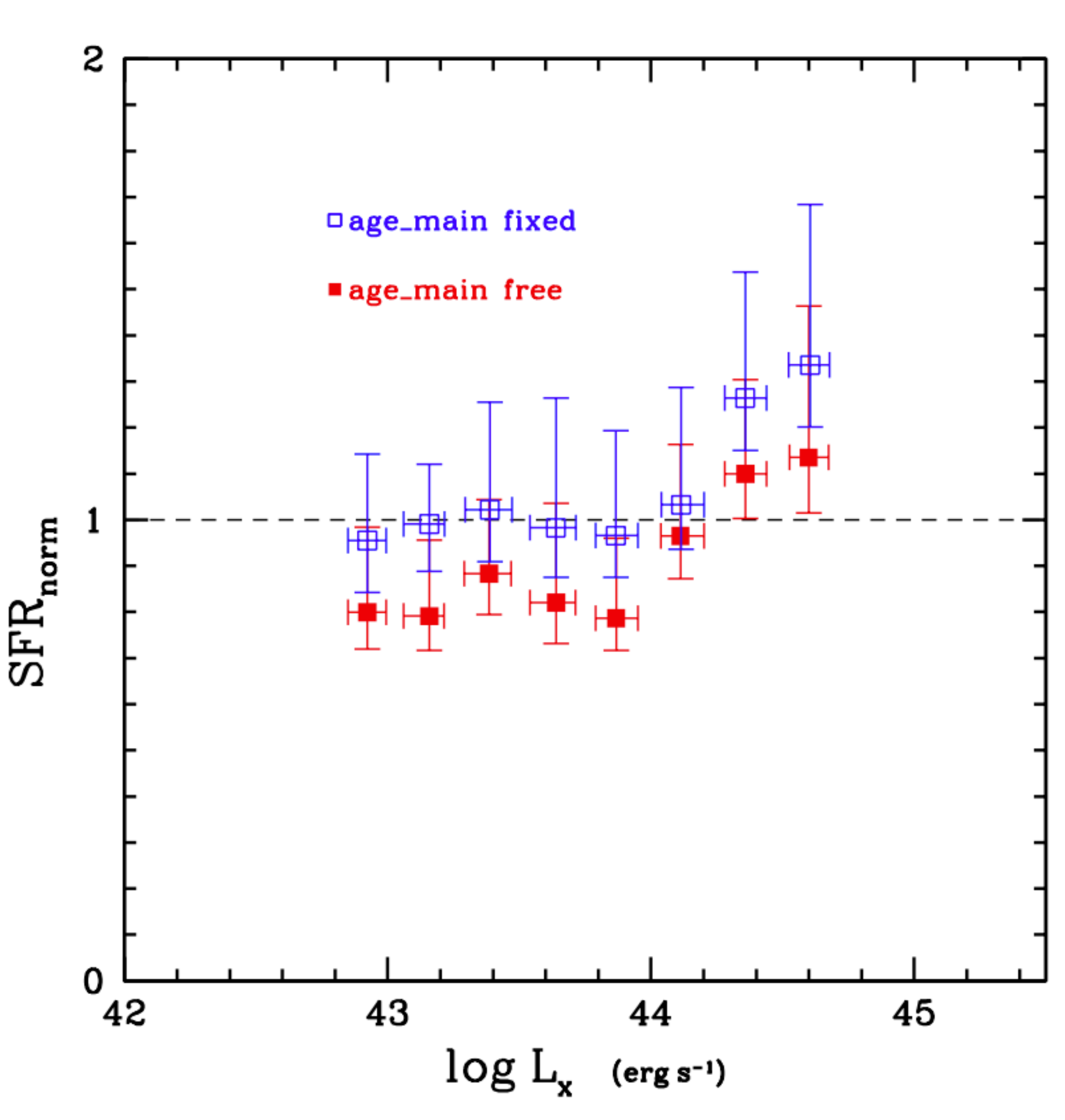}
  \caption{Comparison of SFR$_{norm}$ vs. L$_X$, when the stellar age is free and fixed during the SED fitting process, for the X-ray AGN within $\rm 0.5<z<1.5$. SFR$_{norm}$ remains constant at $\rm L_{X,2-10keV} < 10^{44}\,ergs^{-1}$ and increases at higher L$_X$, in both cases. However, SFR$_{norm}$ values are higher, by on average $\sim 17\%$, when the stellar age is fixed in the SED fitting process. Errors are estimated via bootstrap resampling. The dashed horizontal line indicates the SFR$_{norm}$ value ($=1$) for which the SFR of AGN is equal to the SFR of star-forming galaxies.}
  \label{fig_fixed_sfrnorm}
\end{figure}

In Fig. \ref{fig_fixed_sfrnorm}, we plot the SFR$_{norm}$ vs. L$_X$, for the X-ray AGN within $\rm 0.5<z<1.5$, for the two runs. We notice that the trends are similar, i.e., SFR$_{norm}$ remains constant at $\rm L_{X,2-10keV} < 10^{44}\,ergs^{-1}$ and increases at higher L$_X$. However, SFR$_{norm}$ values are consistently higher, by on average $\sim 17\%$ when the stellar age is fixed in the SED fitting process. Both measurements, though, are in statistical agreement and therefore our overall conclusion does not change, i.e., the SFR of X-ray AGN up to $\rm L_{X,2-10keV} < 10^{44}\,ergs^{-1}$ is consistent with that of star-forming galaxies, while an increase of SFR$_{norm}$ is observed at higher L$_X$.

%\begin{figure}
%\center
%   \begin{tabular}{c c}
%    \includegraphics[height=6cm, width=0.45\textwidth]{sfr_comp_age_free_fixed_agn.pdf} \\
%    \includegraphics[height=6cm, width=0.45\textwidth]{mstar_comp_age_free_fixed_agn.pdf} \\
%    \end{tabular}
%  \caption{Comparison of SFR (top panel) and M$_*$ (bottom panel) measurements with the stellar age free and fixed during the SED fitting process, for the X-ray sample. log\,SFR values are slightly higher (by $\sim 0.02$ on average), when the parameter is free. This trend is more obvious for low SFR values. M$_*$ is lower by, on average, $\sim 0.08$\,dex, when stellar age is set free. As in the case of SFR, this trend is move evident for low M$_*$ values.}
%  \label{fig_fixed_agn}
%\end{figure} 

%\begin{figure}
%\center
%   \begin{tabular}{c c}
%    \includegraphics[height=6cm, width=0.45\textwidth]{sfr_comp_age_free_fixed_reference.pdf} \\
%    \includegraphics[height=6cm, width=0.45\textwidth]{mstar_comp_age_free_fixed_reference.pdf} \\
%    \end{tabular}
%  \caption{Comparison of SFR (top panel) and M$_*$ (bottom panel) measurements with the stellar age free and fixed during the SED fitting process, for the reference cataloug. The results are similar to those for the X-ray sample. However, in this case the trends are more evident.}
%  \label{fig_fixed_reference}
%\end{figure} 

%\section{Reliability of $f_{burst}$ measurements}
%\label{appendix_burst_reliability}

%In this section, we examine the reliability of the f$_{burst}$ parameter. Fig. \ref{fig_fburst_mock} presents the estimated and true values of f$_{burst}$, from the mock analysis. For all three populations, CIGALE can reliable estimate the f$_{burst}$ parameter for values $\leq 0.15$. The algorithm is not sensitive to higher values. In the case of luminous AGN, the code slightly overestimates the parameter for values f$_{burst}<0.10$, but only by 0.017 (median difference between estimated and true values). Compared to the f$_{burst}$ estimates for the sources in the reference galaxy sample, those for the luminous AGN are overestimated by $0.021$ (median value). The offset is lower (0.012) when we compare luminous AGN with their lower L$_X$ counterparts. We conclude that CIGALE can reliably constrain the f$_{burst}$ parameter.

%\begin{figure}
%\centering
%  \includegraphics[width=1.\linewidth, height=7.cm]{fburst_data_mock_all.pdf}
%  \caption{Comparison of the true and estimated values for the $f_{burst}$ parameter, from the mock analysis. CIGALE can reliably estimate the parameter for all the examined populations. The algorithm is not sensitive at $f_{burst}>0.15$. Errors represent the standard deviation of the measurements.}
%  \label{fig_fburst_mock}
%\end{figure} 

\end{document}